# A Comparison of Electric-Field Profile Reconstruction Methods based on the E-FISH Diagnostic

Edwin Setiadi Sugeng[1], Zhijian Yang[1], Yaqi Zhang[1, 2], Anbang Sun[2], Tat Loon Chng[1,*]

[1] Department of Mechanical Engineering, College of Design and Engineering, National University of Singapore, 117575, Singapore
[2] School of Electrical Engineering, Xi'an Jiaotong University, Xi'an, China 710049

*Author to whom any correspondence should be addressed.

E-mail: tatloon@nus.edu.sg

## Abstract

Electric-field induced second harmonic (E-FISH) generation has stood out as a non-intrusive, non-resonant measurement method for electric fields in moderate to atmospheric pressure plasmas. However, the nonlinear, path-integrated nature of E-FISH signal generation remains a principal source of measurement uncertainty, as it precludes a unique relation between a single-point measurement and the corresponding local electric field. One way of addressing this ill-posed problem is to acquire a spatial E-FISH signal profile and invert it to recover the underlying field distribution. While several of these methods have been employed in various studies and differ significantly in terms of their implementation, a rigorous evaluation of their reconstruction accuracy and overall performance has yet to be conducted. This paper aims to compare three such methods, namely machine learning, linear inversion, and Abel inversion. This is facilitated by acquiring E-FISH signals from 4 canonical electrode geometries – sphere, disk, plate and cylinder – which provide electric field profiles with distinct spatial characteristics. These three methods are assessed mainly on their overall performance, robustness to sparse data and noise, and ability to reconstruct asymmetric profiles. All three methods are shown to successfully reconstruct the electric-field profiles, although each presents distinct advantages and limitations. Machine learning offers high reconstruction accuracy and speed together with strong resilience to noise, but its performance deteriorates for profiles that deviate substantially from the distribution of its training dataset, such as strongly asymmetric profiles. The linear inversion method is more versatile and can accommodate a broad range of profile shapes but exhibits sensitivity to the choice of initial profile. Abel inversion is restricted to axisymmetric profiles but performs particularly well for measurements involving horizontal field polarization. These principal findings, particularly the complementary limitations of the machine-learning and linear inversion methods, motivate the conception of an approach for reconstructing electric fields with arbitrary or (a priori) unknown profile shapes. Overall, this study highlights the trade-offs between the three methods and provides a basis for selecting an appropriate reconstruction strategy according to envisaged measurement conditions and characteristics of a given electric-field profile.

## 1. Introduction

The electric field is a quantity that governs fundamental processes in plasmas, such as species generation, recombination, and charge transport. As these processes in turn influence resulting phenomena, for instance fast gas heating and ionic winds, such measurements are crucial for understanding and optimizing plasmas.

Although conventional field probes have been widely used, these are often viewed as invasive and risk distorting the field measurements by perturbing the local electric field. An alternative non-contact method, known as electric-field induced second-harmonic generation (E-FISH), was proposed in 2017 by Dogariu et al. [1]. The underlying optical effect was observed in centrosymmetric media in the mid-1970s, notably by Bigio et al. [2], and has since been successfully applied for measuring electric fields in non-equilibrium plasmas [3-5].

The principle of the E-FISH method is based on detecting the second harmonic light produced by the interaction of a focused laser beam with the externally applied electric field it is propagating through. For a probe beam of power $P^{(\omega)}$, passing through a spatially varying externally applied electric field of $E_{ext}(z)$, the power of the generated second harmonic light ($P^{(2\omega)}$) is given by equation (1) [6, 7]:

$$P^{(2\omega)}(z) \propto \left[\alpha^{(3)} \cdot N \cdot P^{(\omega)}\right]^2 \cdot \frac{1}{z_R} \cdot \left|\int_{-\infty}^{\infty} \frac{E_{ext}(z) e^{i\Delta kz}}{1+i\frac{z}{z_R}} dz\right|^2, \tag{1}$$

where $\alpha^{(3)}$ is the third-order nonlinear hyperpolarizability (a fourth-rank tensor), whose value is dependent on the direction of the electric field relative to the polarization of the probe beam; $N$ is the neutral number gas density; $z$ is the beam propagation axis; $z_R$ is the Rayleigh range of the laser beam; and $\Delta k$ is the wave vector mismatch. It is assumed that the beam focus and the origin of the $E_{ext}(z)$ are both at $z = 0$.

From the above equation, it is observed that the measured signal is dependent on the Gouy phase shift, which manifests itself across the entire electric field distribution along the optical path (through the $1 + i\frac{z}{z_R}$ term) [6], rather than merely the external field within the focal region. Thus, an unknown field profile shape (inadvertently) introduces uncertainty into a single-point E-FISH measurement.

To address this problem, two solutions have been identified within the community. The first, as discussed in [8-11], relies on using two laser beams crossed at a fixed angle to constrain the interaction region. This, however, results in a trade-off between signal strength and spatial resolution, prompting further research into strategies for signal amplification [11]. The second, which is relevant to this work, relies on displacing the electric field profile with respect to the laser beam focus to acquire an E-FISH signal profile. This signal profile is subsequently inverted to reconstruct the shape of the unknown electric field distribution. Several reconstruction methods have since been proposed, such as the use of machine-learning models [12-14], an inverse-Abel-type transform [15, 16], as well as linear inversion [17-19].

Machine learning models trained for E-FISH inversion are generally robust, and have proven resilient even for noisy and sparse data [20, 21]. However, they are broadly restricted to a certain set of parameters and profiles included in their training dataset. Most notably, the lack of a systematic way to produce synthetic data relevant to asymmetric profiles has limited its applicability to symmetric field distributions. On the other hand, linear inversion iteratively converges on a solution by progressively refining upon an initial guess, analogous to solving an equation for an unknown (scalar) root, but extended in this case, to an unknown *profile.* As such, this approach, though in principle not restricted to any profile type, is often very sensitive to noise and the initial values input to the model [17-19]. Finally, Abel inversion, as is well-known in the community, applies only to axisymmetric field distributions [15, 16, 22], but entails the sampling of less information (e.g. half a profile), thereby ostensibly accelerating the data acquisition process.

Although the above offers brief insight into the relative performance of each method, a direct comparison of these three methods (in reconstructing electric fields) remains missing. As such, this work aims to *experimentally* compare the accuracy and robustness of the various inversion methods in reconstructing an underlying electric profile.

The sections that immediately follow first discuss the experimental setup and the various electrode geometries used. Thereafter, the acquired experimental data and reconstruction results are presented and discussed.

## 2. Methodology

### 2.1. E-FISH Optical Setup

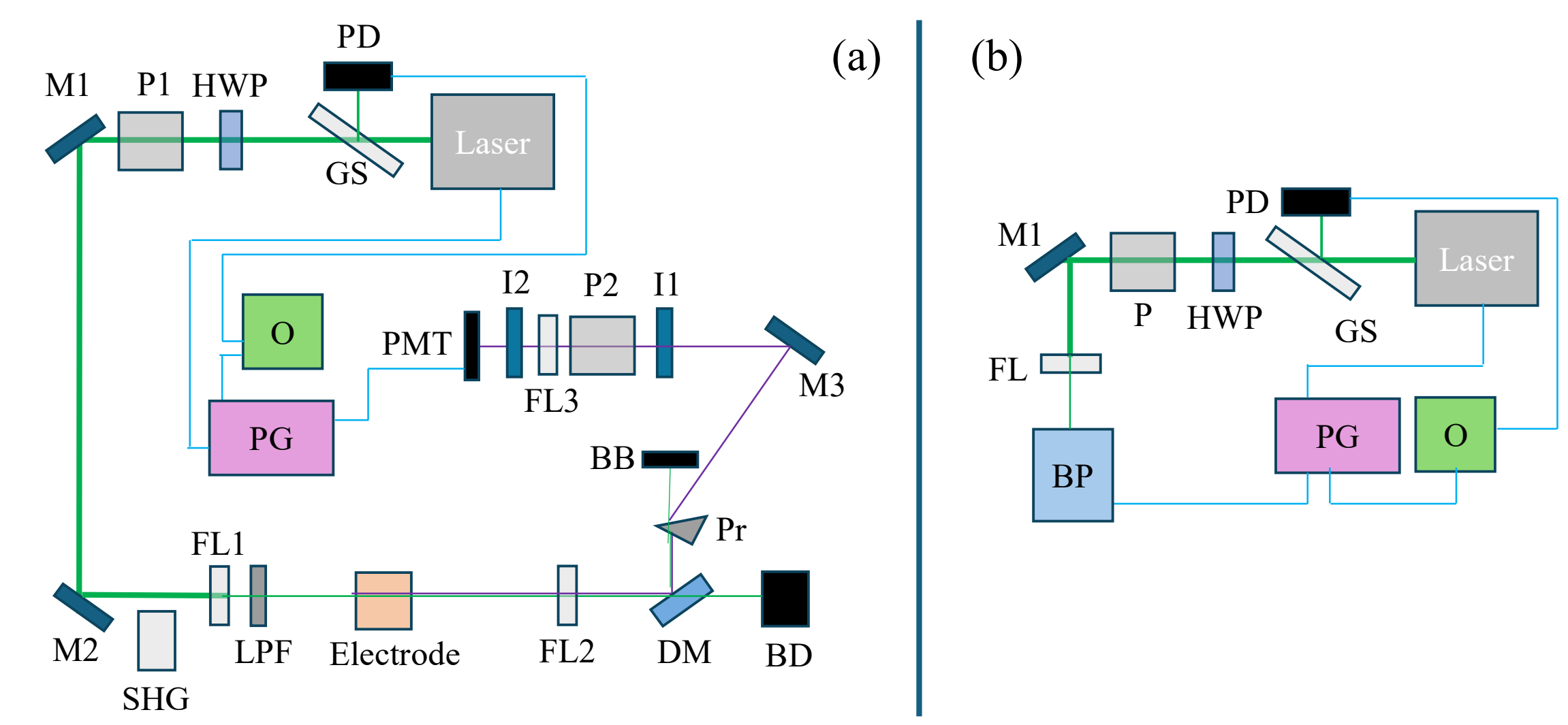

**Figure 1**. (a) Schematic of the E-FISH experimental setup and (b) beam profiling setup. PD: photodiode; GS: glass slide; HWP: half-wave plate; P: polarizing beamsplitter; M: mirror; SHG: second harmonic generating crystal; FL: focusing lens; LPF: long-pass filter; DM: dichroic mirror; BD: beam dump; BB: beam block; Pr: prism; I: iris; PMT: photomultiplier tube; O: oscilloscope; PG: pulse delay generator; BP: beam profiler. The SHG is only placed along the beam path during alignment.

Figure 1(a) shows the setup used to acquire the E-FISH signal produced by a variety of electrode geometries listed in Table 1. The beam is delivered via an EKSPLA PL-2251A laser system with a nominally 26 ps pulse width and maximum pulse energy of 50 mJ at 1064 nm wavelength. Approximately 3 mJ of energy per pulse is used for this study. The beam is first passed through a glass slide, reflecting about 4% into a photodiode (PD; Thorlabs DET10A2) to monitor the fundamental beam intensity. Thereafter, the beam is let through a half-wave plate (HWP; Thorlabs WPH10M-1064) and polarizer (P1; Thorlabs CCM1-PBS25-1064-HP/M) for energy and polarization control, before being reflected by a pair of high reflectivity mirrors (M1, M2; Thorlabs PF20-03-F01). To maximize signal, the beam polarization is aligned parallel to the electric field component being probed. A long pass filter (LPF; Thorlabs FGL850S) is used to eliminate any stray second harmonic light generated by the upstream birefringent optics (especially the HWP). A second harmonic crystal (SHG) is placed before FL1 to generate a weak second harmonic beam for alignment purposes and is removed when experiments are conducted. The beam is then focused using a plano-convex lens (FL1; Thorlabs LA4538-ML or Thorlabs LA4782-ML) to the measurement location (see Table 1 and Table 3 for more details). The fundamental and second harmonic E-FISH signal are then collimated using another plano-convex lens (FL2) with the same focal length as FL1. These two beams are subsequently separated: first using a dichroic mirror (DM; Thorlabs HBSY22), followed by a prism (Pr; Thorlabs PS704). The fundamental beam is directed towards the beam dump (BD) and beam block (BB), while the isolated second harmonic E-FISH signal is directed towards a photomultiplier tube (PMT; Hamamatsu H11706-40) using a high reflectivity mirror (M3). Prior to entering the PMT, the signal first passes through a polarizer (P2; Thorlabs CCM1-PBS25-532-HP/M) to probe only the polarization component parallel to the electric field component, before being focused via a 10-cm plano-convex lens (FL3; Thorlabs LA4545-ML) to increase its intensity. A pair of irises (I1, I2) serve as spatial filters to reject any remaining stray light. The PMT and Laser are synchronized using a pulse delay generator (PG; Stanford Research Systems DG-545). The signals from the PD and the PMT are acquired using an oscilloscope (O; Teledyne Test Tools T3DSO31004).

### 2.2. Measurements of Rayleigh Range

As the Rayleigh range critically affects the E-FISH signal (see equations (1) and (3)), a beam profiler (BP; Thorlabs BC207VIS/M) is used to estimate the Rayleigh range for the experimental setup discussed previously in Figure 1(a). The beam profiler is installed according to Figure 1(b).

In Figure 1(b) (which shares many similarities with the E-FISH setup), the 1064 nm probe beam is passed through a glass slide (GS), of which a fraction is picked off and directed towards a photodiode (PD; Thorlabs DET10A2). The PD is connected to the oscilloscope (O), and the pulse generator (PG; Berkeley Nucleonics Corporation Model 577) is used to synchronize the BP and Laser via the pulse generator. The transmitted portion of the probe beam passes through a half-wave plate (HWP) and polarizer (P), used to adjust the intensity of the laser beam arriving at the BP. Mirror (M1) directs the beam towards the focusing lens (FL), a 25-cm or 50-cm lens, depending on the Rayleigh range to be measured. This lens (FL) is mounted on a rail (Thorlabs RLA300/M, RC2/M; identical to the one in Figure 2), allowing the beam profile to be captured (using the BP) at various locations along its optical path.

A typical beam profile measurement involves a $\pm 100$ mm scan from the focus, subsequently fitted to obtain the divergence angle. Using this divergence angle, the beam waist and Rayleigh range can be inferred. From these calculations, the Rayleigh range for the 1064 nm beam with a 25 cm lens lies between 1.4 to 6.7 mm and with a 50 cm lens is approximately 5.8 to 12.7 mm. These values are used as an input to the various inversion models, along with a constant wave vector mismatch ($\Delta k \propto n^{(\omega)} - n^{(2\omega)}$, with $n$ being the refractive index at the given wavelength) of $-48.3\ m^{-1}$ (see section 2.4).

### 2.3. E-FISH Signal Acquisition

Seven different electrode geometries are chosen in this experiment (see details in Table 1), namely a 10 mm diameter sphere; 10 mm cylinders; 20 x 50-mm and 30 x 50-mm plates; 5-mm and 10-mm diameter disks; as well as a surface dielectric barrier discharge (SDBD). The vertical electric field component of the first six geometries primarily generates field profiles of differing full-widths-at-half-maximum (FWHM), with the plates producing (almost) flat-top profiles. In contrast, the horizontal electric field component of these geometries yields field profiles that are antisymmetric about $z$ = 0 (beam focus), with each of the two peaks located equidistant away. Also of note are tests on the dielectric barrier discharge, whose *true* electric field shape remains unknown, and are used primarily to assess the performance of the various inversion models on actual discharge data.

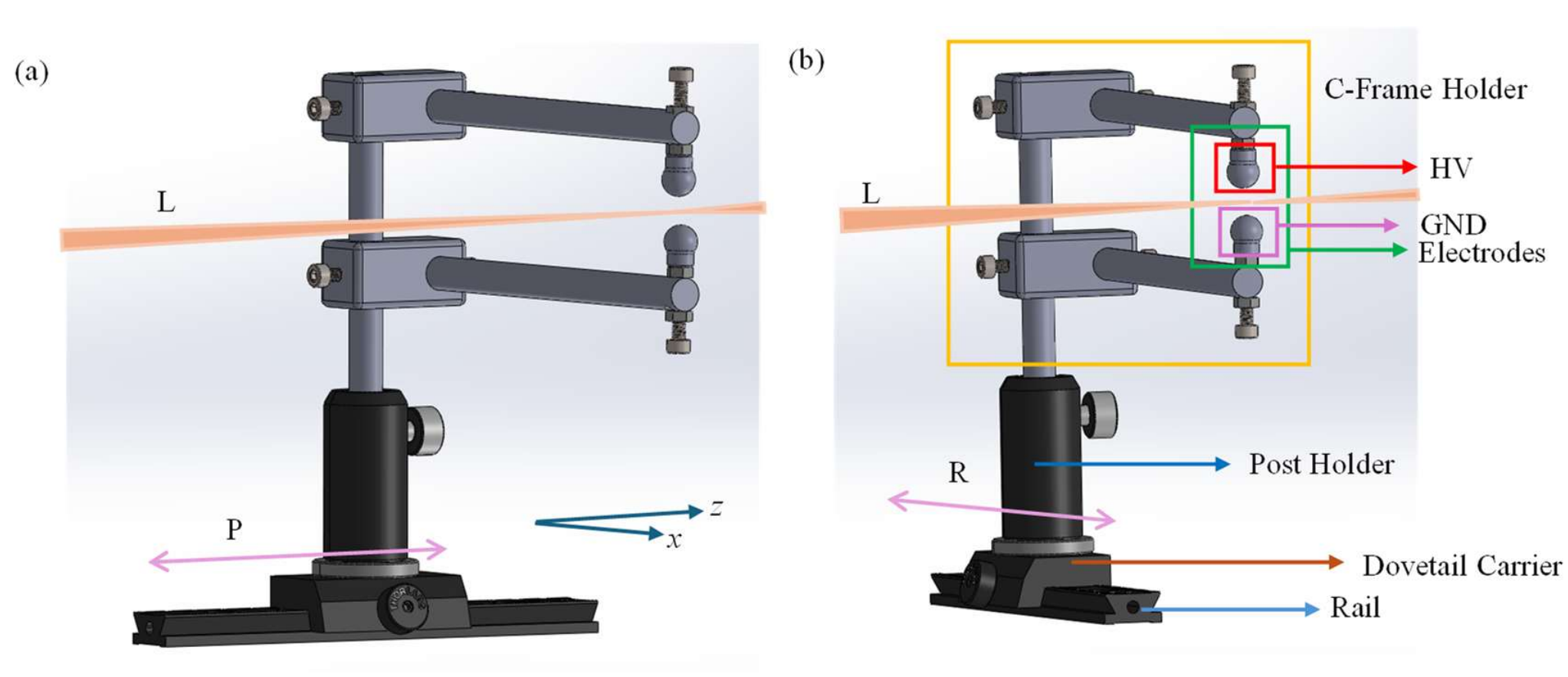

**Figure 2**. Schematic of the different configurations for (a) parallel scanning, indicated by arrow 'P'; and (b) perpendicular (radial) scanning, indicated by arrow 'R'. The laser beam is indicated as L in both schematics. Although spherical electrodes are shown for illustrative purposes, the same mounting arrangement is used for the disk and plate electrodes (not shown). $z$ refers to the direction of the $z$-axis, which is parallel to the beam (L) propagation direction; and $x$ refers to the radial direction, parallel to R.

**Table 1**. List of electrode geometries used in the experiment and the data collection range. HV refers to the high voltage electrode; GND refers to the ground electrode; DI refers to the dielectric layer; $H$ refers to the teeth height; $L$ refers to length; $W$ refers to width; $D$ refers to the diameter; $G$ refers to the interelectrode gap; $T$ refers to thickness; and $z_R$ refers to Rayleigh range (see also Table 3 for more details on the Rayleigh ranges tested). $z$ refers to the beam propagation direction (see equation (1)) and $x$ refers to the radial direction (see equation (3)).

<table>
<tr><th>Electrode</th><th>Sphere</th><th>Disk</th><th>Plate</th><th>4 Cylinders</th><th>SDBD</th></tr>
<tr><td>Diagram</td><td>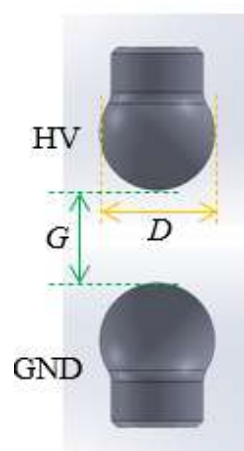
</td><td>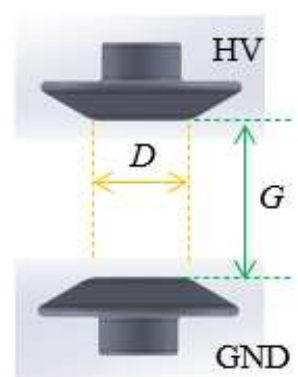
</td><td>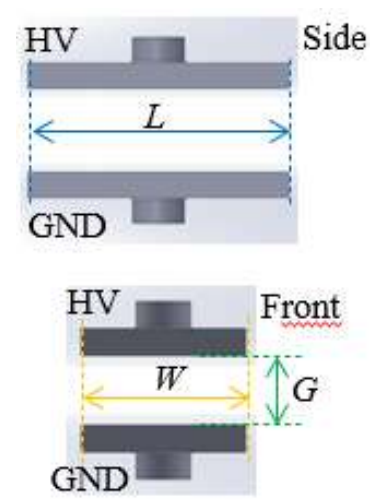
</td><td>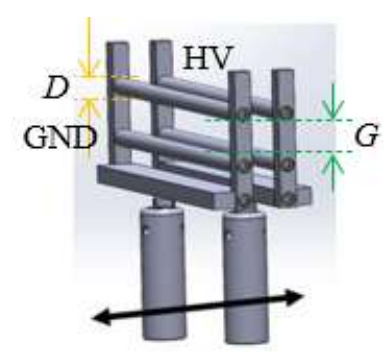
</td><td>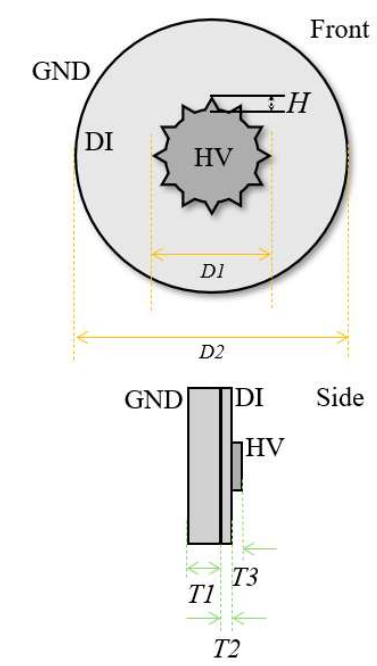
</td></tr>
<tr><td>Dimensions [mm]</td><td>$D$: 10<br>$G$: 10</td><td>$D$: 5, 10<br>$G$: 10</td><td>$L$: 50<br>$W$: 20, 30<br>$G$: 10</td><td>$D$: 10<br>$G$: 10</td><td>$H$: 2<br>$T1$: 0.05<br>$T2$: 1<br>$T3$: 0.3<br>$D1$: 20; $D2$: 60</td></tr>
<tr><td>Lens Focal Length [cm]</td><td colspan="4">25, 50</td><td>25</td></tr>
<tr><td>Measurement Location</td><td colspan="4">Vertical Polarization: middle height (half $G$)<br>Horizontal Polarization: 3 mm below the HV tip</td><td>2 mm from HV<br>2 mm from DI</td></tr>
<tr><td rowspan="8">Data Collection Range and Grid (laser focus at $z = 0$ mm)</td><td colspan="4">Parallel scanning direction, 25 cm focal length (smaller $z_R$ value)</td><td rowspan="8">$z = [-30,30]$ mm</td></tr>
<tr><td colspan="4">$z = [-75,75]$ mm, total of 71 points<br>$\Delta z = 10$ mm for [-75, -55] mm and [55, 75] mm<br>$\Delta z = 5$ mm for (-55, -40] mm and [40, 55) mm<br>$\Delta z = 2$ mm for (-40, -20] and [20, 40] mm<br>$\Delta z = 1$ mm for (-20, 20) mm</td></tr>
<tr><td colspan="4">Parallel scanning direction, 50 cm focal length (larger $z_R$ value)</td></tr>
<tr><td colspan="4">$z = [-230,230]$ mm, total of 147 points<br>$\Delta z = 10$ mm for [-230, -120] mm and [120, 230] mm<br>$\Delta z = 5$ mm for (-120, -60] mm and [60, 120) mm<br>$\Delta z = 2$ mm for (-60, -40] mm and [40, 60) mm<br>$\Delta z = 1$ mm for (-40, 40) mm</td></tr>
<tr><td colspan="4">Radial scanning direction, 25 cm focal length (smaller $z_R$ value)</td></tr>
<tr><td colspan="4">$x = [-60,0]$ mm, total of 35 points<br>$\Delta x = 5$ mm for [-60, -35] mm<br>$\Delta x = 2$ mm for (-35, -15] mm<br>$\Delta x = 1$ mm for (-14, 0) mm<br><br>Only for sphere, $x = [-25,0]$ mm, total of 25 points, $\Delta x = 1$ mm</td></tr>
<tr><td colspan="4">Radial scanning direction, 50 cm focal length (larger $z_R$ value)</td></tr>
<tr><td colspan="4">$x = [-60,0]$ mm, total of 35 points<br>$\Delta x = 5$ mm for [-60, -40] mm<br>$\Delta x = 2$ mm for (-40, -20] mm<br>$\Delta x = 1$ mm for (-20, 0) mm</td></tr>
</table>

The resulting variety of profiles encompassed by the different Rayleigh ranges and electrode geometries is deemed sufficient to provide a meaningful comparison of the three reconstruction methods.

To acquire the E-FISH (signal) profile, the spherical, disk, and plate electrodes are mounted on a C-Frame Holder, as shown in Figure 2. One of the electrodes is designated as the high voltage (HV; 5 kV for the vertical polarization and up to 16 kV for the horizontal polarization) and the other is the ground (GND). A high voltage direct-current power supply (Glassman High Voltage Inc.) can deliver a peak voltage of up to 20 kV respectively.

The C-Frame Holder is inserted into a Post Holder and mounted onto a rail with a dovetail carrier (Thorlabs RLA300/M, RC2/M, and RC3) as mentioned above. This setup allows a scanning resolution of up to 1 mm both parallel (see arrow P in Figure 2(a)) and perpendicular (see arrow R in Figure 2(b); termed henceforth as the radial direction) to the optical path. This also enables the examination of different inversion methods that require scanning over distinct directions (i.e., parallel – machine learning and linear inversion; perpendicular – Abel transform). The sampling range and grid are listed in Table 1, typically terminating where the signal at the wings drops to about 10% of its peak value. This threshold was chosen as it almost reaches the baseline noise level in these experiments.

Due to their two-dimensional geometry, the cylindrical electrodes are mounted on a different frame as shown in Table 1. Both frames are connected to a post, followed by a Post Holder. The Post Holder is also mounted onto the identical Rail as in the other experiments. Further details of the cylinder experimental setup are similar to that found in Yang et al. [12].

The dielectric barrier discharge is driven by a nanosecond high-voltage pulsed power supply (FID Technologies GmbH, FPG 15NS-15) and mounted on translation stages to enable displacement of the electrode position relative to the laser beam.

## 2.4. Details of Electric Field Reconstruction Methods

As mentioned earlier, three different reconstruction methods, namely machine learning [13], linear inversion [17, 18], and Abel inversion [15, 16] are applied to the acquired E-FISH profiles.

### 2.4.1. Machine Learning

The machine learning model employed in this study is the Polarization-Conditioned Fourier-enhanced DeepONet (PC-FDON) developed by Yang et al. [13], a model capable of reconstructing electric fields from both vertically and horizontally polarized E-FISH signals. Compared with our previous Decoder-DeepONet (DDON) [14] specifically designed for vertically polarized E-FISH signals, the PC-FDON incorporates a Fourier sub-branch that captures the frequency component arising from the Gouy phase shift, together with a dedicated polarization branch and gated unit that tailors the predictions to each polarization state. The model has been trained on 1,147,147 profile pairs (i.e., electric field and its corresponding E-FISH signal), composed in equal proportions of horizontally polarized E-FISH signals (with $E$ adopting a bipolar shape) and vertically polarized E-FISH signals (with $E$ adopting a bell-shaped or double-peaked profile). Furthermore, by training over a wider range of phase-mismatch values ($u = \Delta k \cdot z_R$) and incorporating this parameter as an input, the PC-FDON is expected to accurately reconstruct electric fields from E-FISH measurements acquired across different values of the wave vector mismatch ($\Delta k$) and Rayleigh range ($z_R$).

### 2.4.2. Linear Inversion

The linear inversion method reconstructs an (unknown) electric field profile by minimizing the mean square error (MSE) between the experimentally acquired E-FISH signal and the E-FISH signal obtained by integrating the reconstructed electric field following equation (1).

The spatial z-locations of the experimental data points comprising each profile are retained and reused for the reconstruction process. The initial guess is generated using Lorentzian curves of varying widths, or a combination thereof, depending on the Laplacian field profile for the given electrode geometry. Lorentzian curves (and their combinations) are chosen as they are representative of canonical electric fields [12]. It can be represented by equation (2) below.

$$E(z) = \frac{1}{1 + 4\left(\frac{z - c}{a}\right)^2}$$

(2)

On equation (2), $z$ represents the direction of beam propagation, as also used in equation (1), $c$ represents the distance of the peak of the profile from $z$ = 0, and $a$ represents the full width half maximum (FWHM) of the profile. A single-peak Lorentzian is used for the vertically polarized fields, while a bipolar profile comprising two equal-width Lorentzian curves with non-zero value of $c$ is used for the horizontal polarization. The Laplacian field is simulated using the MATLAB Partial Differential Equation (PDE) toolbox. Along with bounds of [0,1] and [-1,1] for the vertical and horizontal polarization respectively, the initial guess is used as the input to the *lsqnonlin* function in MATLAB.

In addition to minimising the MSE between the experimentally acquired E-FISH signal and integration of the reconstructed electric field, the matrices are appended with the first and second derivatives of the reconstructed electric field. The weighting parameters for the first and second derivatives are swept, using a logarithmic scale between $10^{-6}$ to $10^{2}$. The combination of the initial guess and weighting parameters giving rise to the lowest MSE values between the Laplacian and reconstructed electric fields is chosen for this study (see Appendices A and B).

### 2.4.3. Abel Inversion

The Abel inversion reconstructs the original field profile from a signal generated by axisymmetric geometries (i.e. sphere and disks). As opposed to the first two methods, the radial distribution of the E-FISH signal is required as an input for the reconstruction, thereby necessitating a different scanning direction. The E-FISH signal at a transverse (radial) offset $x$ (see Figure 2) is given by equation (3).

$$P^{(2\omega)}(x) \propto \left[\alpha^{(3)} \cdot N \cdot P^{(\omega)}\right]^2 \cdot \frac{1}{z_R} \cdot \left| \int_{-L}^{L} \frac{E_{ext}(r) e^{i\Delta kz}}{1 + i\frac{z}{z_R}} dz \right|^2$$

$$r = \sqrt{x^2 + z^2}$$

(3)

In reconstructing the electric field, a weight matrix, $W$, which encompasses the Gouy phase shift terms, is used as the factor relating the electric field ($E$) to the E-FISH signal ($P$) (see equation (4)). The smoothing parameter $c$, based on the second derivative of the electric field is also applied, to reduce its sensitivity to noise.

$$|W \cdot E + c \cdot E''| = |P|$$

(4)

The objective is therefore to find $E$ that minimizes the error between the left- and right-hand sides of equation (4). The smoothing parameters are chosen by sweeping over a range of values, from 0 to 1. The smoothing parameter that gives the lowest MSE is then selected for that profile, as listed in Appendices A and B. The implementation details closely follow the derivation provided by the work of Guo et al. [15].

## 3. Results and Discussion

The implementation of the three methods is first validated using synthetically generated inputs. Thereafter, using experimental data, the three methods are evaluated in terms of their accuracy by comparing their MSE. The sensitivity of each method to sparse data conditions; the input Rayleigh range; initial guess; and data noise are then compared. Beyond this analysis, the models are further assessed for their performance on asymmetric electric field profiles, followed by a discussion on the handling of such cases, as well as instances where the underlying electric field profiles are unknown. The final section is devoted to a summary and comparison of the methods' practical aspects of implementation.

### 3.1. Method Validation

**Table 2**. Parameters of the Lorentzian profile and Rayleigh range used to validate the three reconstruction methods. The parameters $a$ and $c$ correspond to that used in equation (2). The profiles are integrated according to equation (1) using the Rayleigh range to generate the simulated E-FISH signal. The resulting MSEs between the reconstructed and actual electric fields for the machine learning method, linear inversion method, and Abel inversion method are also shown as MSE ML, MSE Lin, and MSE Abel, respectively.

| # | Lorentzian Profile | | Rayleigh Range [m] | MSE ML | MSE Lin | MSE Abel |
|---|---|---|---|---|---|---|
| | $a$ [m] | $c$ [m] | | | | |
| V-1 | 0.006 | 0 | 0.003 | 5.76E-4 | 1.52E-3 | 9.24E-5 |
| V-2 | 0.02 | 0 | 0.003 | 1.21E-4 | 1.44E-3 | 1.44E-3 |
| V-3 | 0.035 | 0 | 0.003 | 9.00E-4 | 2.50E-4 | 7.38E-3 |
| V-4 | 0.006 | $\pm$0.006 | 0.003 | 4.58E-2 | 1.20E-1 | 1.49E-4 |

As an initial validation check, the performance and implementation of the three reconstruction methods are evaluated using synthetic Lorentzian profiles, given by equation (2). The parameters used to generate the profiles for this validation are listed in Table 2. Their reconstruction capability is examined by visually observing the fit between the reconstructed electric field profiles (see Figure 3) and their true distributions as well as by inspecting their MSEs (see Table 2).

The generally low MSE values ($\leq$ 1.5E-3 with the exception of V-4), and the good visual agreement with these synthetically prescribed profiles, validate our underlying algorithms for the three reconstruction methods, providing a firm basis for their subsequent evaluation using experimental data. The elevated MSE for the linear inversion for the bipolar case arises due to the non-uniqueness of the profiles, which will be further discussed in the following sections.

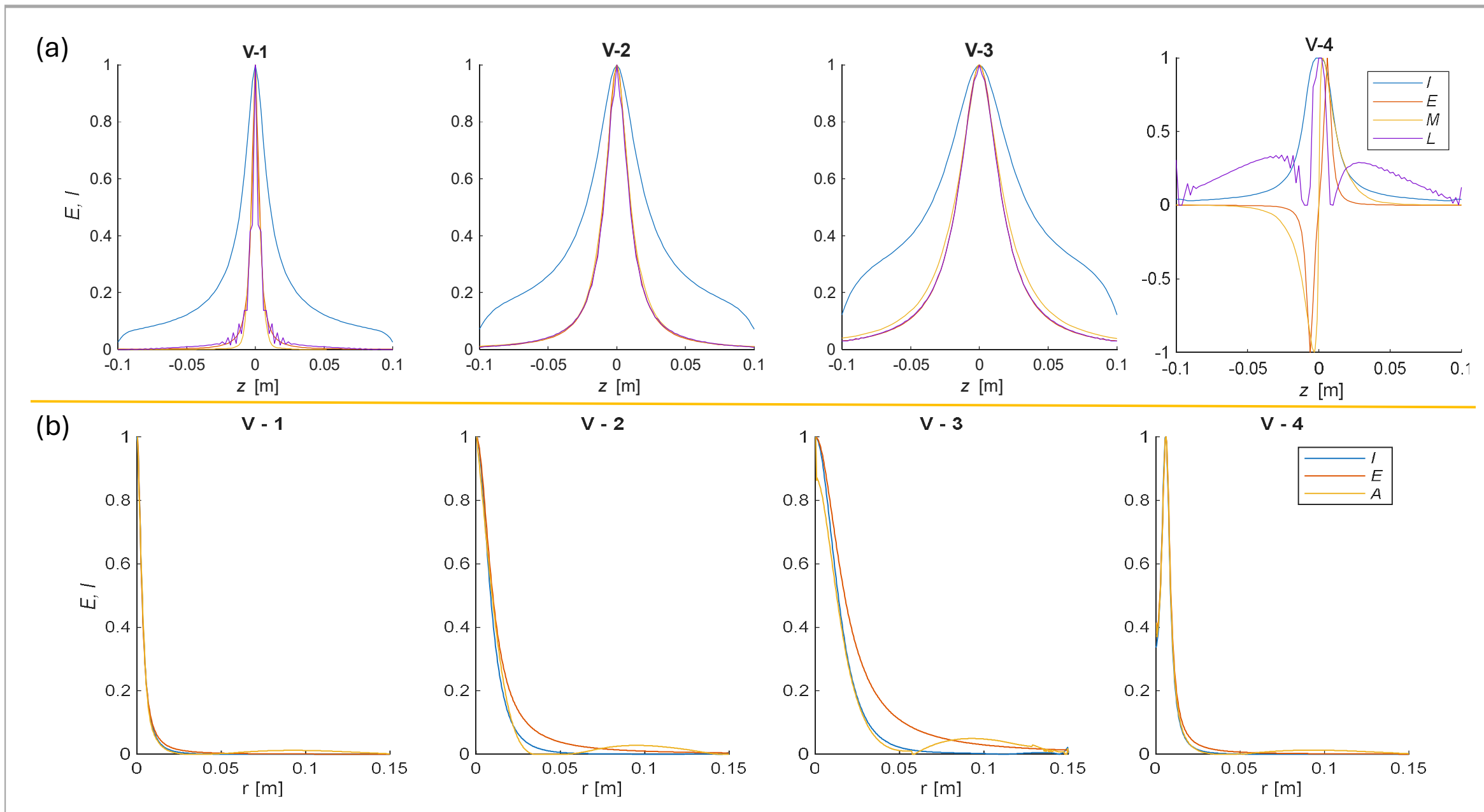

**Figure 3**. Validation cases for (a) machine learning ($M$) and linear inversion ($L$) methods and (b) for Abel inversion ($A$). $E$ shows the electric field and $I$ the simulated E-FISH signal, generated by integrating $E$ according to equation (1) for plot (a) and equation (3) for plot (b).

**Table 3**. List of electrode geometries and Rayleigh ranges for electric field reconstruction and the methods evaluated. For more details on the electrode appearance and data collection, refer to Table 1.

| # | Polarization | Method | Electrode | Dimension [mm] | $z_R$ [m] |
|---|---|---|---|---|---|
| S-V-1 | Vertical | Machine Learning; Linear Inversion | Sphere | *D*: 10 | 0.00582 |
| S-V-2 | Vertical | Machine Learning; Linear Inversion | Sphere | *D*: 10 | 0.00752 |
| D1-V-1 | Vertical | Machine Learning; Linear Inversion | Disk | *D*: 5 | 0.00146 |
| D1-V-2 | Vertical | Machine Learning; Linear Inversion | Disk | *D*: 5 | 0.00582 |
| D2-V-1 | Vertical | Machine Learning; Linear Inversion | Disk | *D*: 10 | 0.00496 |
| D2-V-2 | Vertical | Machine Learning; Linear Inversion | Disk | *D*: 10 | 0.01272 |
| P1-V-1 | Vertical | Machine Learning; Linear Inversion | Plate | *W*: 20 *L*: 50 | 0.00146 |
| P1-V-2 | Vertical | Machine Learning; Linear Inversion | Plate | *W*: 20 *L*: 50 | 0.01262 |
| P2-V-1 | Vertical | Machine Learning; Linear Inversion | Plate | *W*: 30 *L*: 50 | 0.00186 |
| P2-V-2 | Vertical | Machine Learning; Linear Inversion | Plate | *W*: 30 *L*: 50 | 0.00982 |
| C-V-1 | Vertical | Machine Learning; Linear Inversion | Cylinder | *D*: 10 | 0.00672 |
| C-V-2 | Vertical | Machine Learning; Linear Inversion | Cylinder | *D*: 10 | 0.00842 |
| S-Va-1 | Vertical | Abel Inversion | Sphere | *D*: 10 | 0.00146 |
| S-Va-2 | Vertical | Abel Inversion | Sphere | *D*: 10 | 0.00582 |
| D1-Va-1 | Vertical | Abel Inversion | Disk | *D*: 5 | 0.00146 |
| D1-Va-2 | Vertical | Abel Inversion | Disk | *D*: 5 | 0.00582 |
| D2-Va-1 | Vertical | Abel Inversion | Disk | *D*: 10 | 0.00146 |
| D2-Va-2 | Vertical | Abel Inversion | Disk | *D*: 10 | 0.00582 |
| S-H-1 | Horizontal | Machine Learning, Linear Inversion | Sphere | *D*: 10 | 0.00146 |
| S-H-2 | Horizontal | Machine Learning, Linear Inversion | Sphere | *D*: 10 | 0.00582 |
| D1-H-1 | Horizontal | Machine Learning, Linear Inversion | Disk | *D*: 5 | 0.00146 |
| D1-H-2 | Horizontal | Machine Learning, Linear Inversion | Disk | *D*: 5 | 0.00582 |
| D2-H-1 | Horizontal | Machine Learning, Linear Inversion | Disk | *D*: 10 | 0.00146 |
| D2-H-2 | Horizontal | Machine Learning, Linear Inversion | Disk | *D*: 10 | 0.00582 |
| S-Ha-1 | Horizontal | Abel Inversion | Sphere | *D*: 10 | 0.00146 |
| S-Ha-2 | Horizontal | Abel Inversion | Sphere | *D*: 10 | 0.00582 |
| D1-Ha-1 | Horizontal | Abel Inversion | Disk | *D*: 5 | 0.00146 |
| D1-Ha-2 | Horizontal | Abel Inversion | Disk | *D*: 5 | 0.00582 |
| D2-Ha-1 | Horizontal | Abel Inversion | Disk | *D*: 10 | 0.00146 |
| D2-Ha-2 | Horizontal | Abel Inversion | Disk | *D*: 10 | 0.00582 |

### 3.2. Method Accuracy

The reconstruction results of the experimental E-FISH signals (*Experimental*) listed in Table 1 and Table 3 are shown in Figure 4 (for the machine learning and linear inversion method) and Figure 5 (for the Abel inversion method).

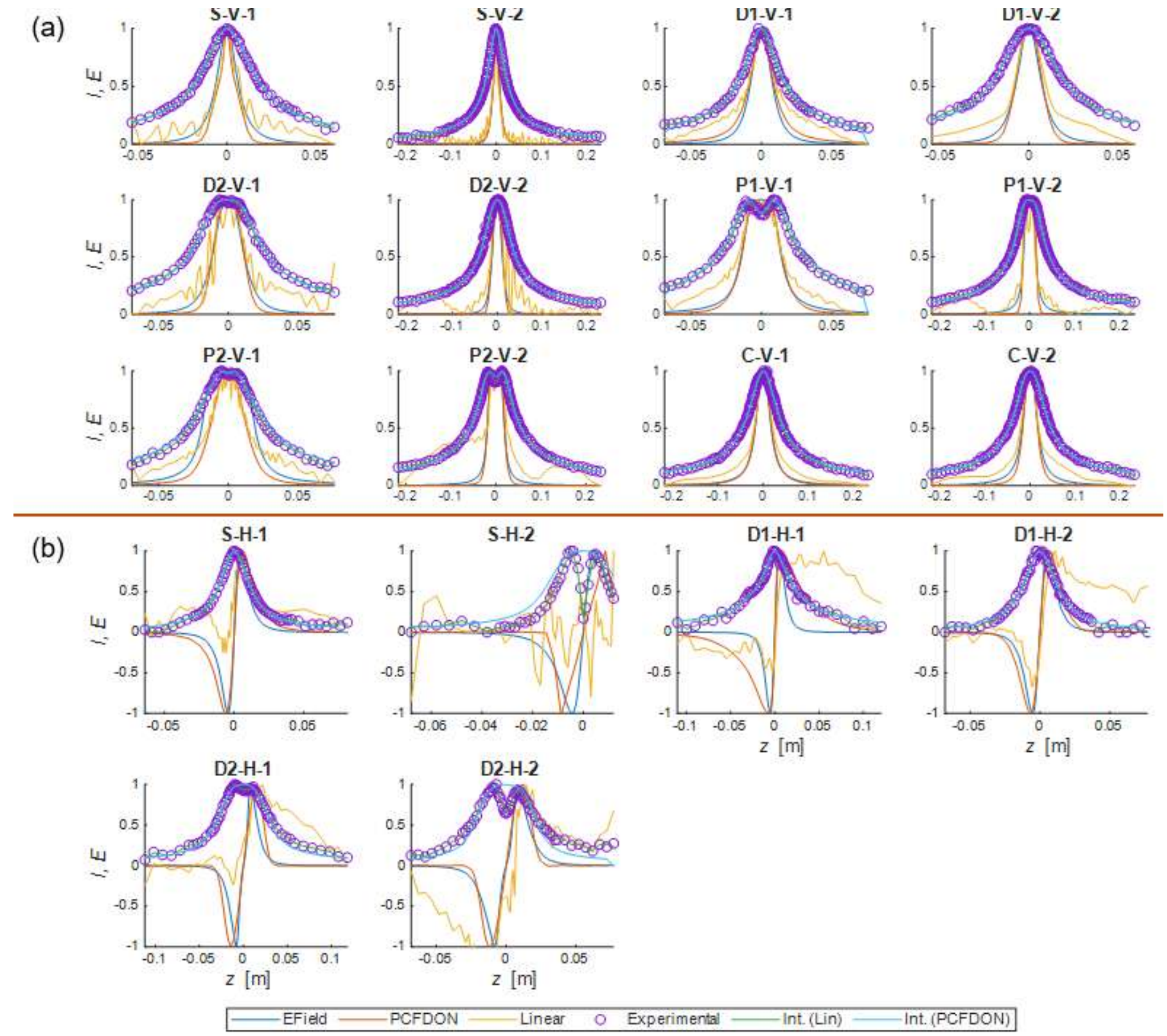

**Figure 4**. Reconstruction results for the machine learning method (PC-FDON) and linear inversion method (Linear) for the (a) vertical polarization and (b) horizontal polarization. Int. (Lin) represents the forward E-FISH obtained from the reconstruction using the linear inversion (Linear) and Int. (PCFDON) is that obtained from the machine learning method (PCFDON). Each subplot title follows the numbering on the first column of Table 3. *I* refers to the normalized E-FISH signal intensity for the Experimental; *E* refers to the electric field normalized values.

The MSE for each of the profiles is shown in Figure 6 for both the vertical and horizontal field components and is obtained by comparing each component against its corresponding Laplacian profile from numerical simulations. In the immediate discussion below, these values are termed as the *baseline* MSE in reference to Figure 6 shown later.

Considering the vertical polarization, the *baseline* MSE values of ≤ 2.59E-3 across the 12 different profile types (shown in Figure 6(a)) indicate that the machine learning model is able to accurately reconstruct the electric field. This good performance is broadly attributable to the good overlap between the tested field profiles and those of the model's training dataset. Reconstruction of profiles that lie entirely outside of the training family (e.g. asymmetric distributions) is likelier to result in a poorer fit, as also evident in section 3.4. It is also important to re-emphasize here that the inputs to the machine learning model are pre-classified into vertically and horizontally polarized categories. This further enhances its performance by allowing the algorithm to effectively distinguish between bell-shaped and antisymmetric profiles for vertical and horizontal electric fields which may both result in a similar E-FISH spatial distribution.

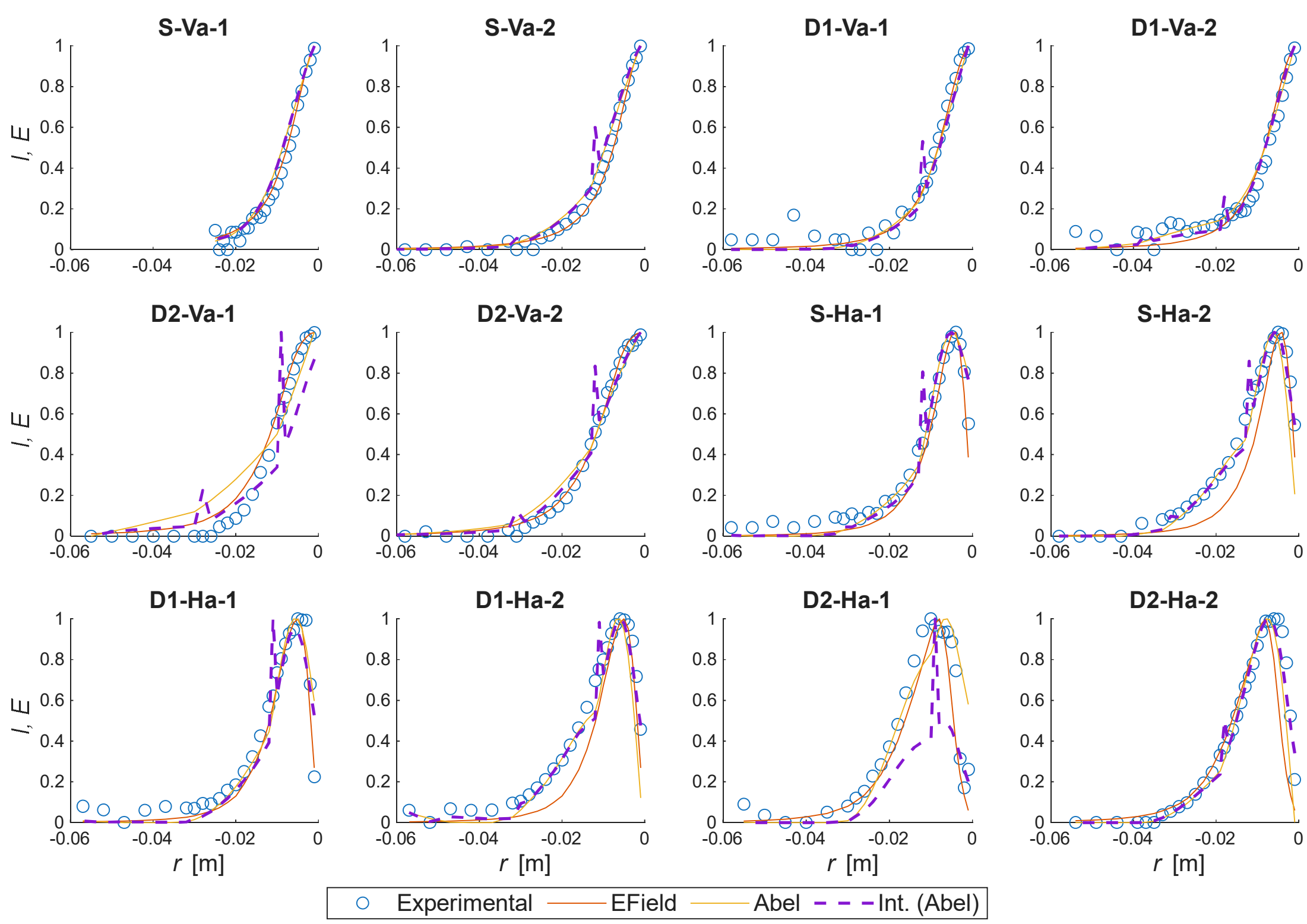

**Figure 5**. Reconstruction results for the Abel inversion method. Each subplot title follows the numbering on the first column of Table 3. Int. (Abel) represents the forward E-FISH obtained from the reconstruction using the Abel inversion (Abel). *I* refers to the normalized E-FISH signal intensity for the Experimental; *E* refers to the electric field normalized values.

The linear inversion method is also able to reconstruct vertical polarization profiles with reasonable accuracy, yielding a *baseline* MSE of ≤ 4.28E-2 for all profiles. As this method relies solely on minimizing the error between the experimentally acquired data and the forward E-FISH (i.e. the residual between the reconstructed profile and equation (1); see lines Int. (Lin) and Int. (PCFDON) in Figure 4), an optimized solution may be obtained even with an electric field profile different from the *true* Laplacian field. This especially applies if no constraints are placed on the symmetry of the output profile. In such cases, the resulting solution is often non-unique and can simply be accommodated by favouring one section of a given distribution over another. For example, in D2-V-2 of Figure 4(a), a slightly narrower peak region can be accordingly compensated with elevated values near the wings. This is one of the possible reasons for the generally higher MSE.

Finally, the Abel inversion approach exhibits a similar level of accuracy to that of machine learning. Figure 6(a) shows that the reconstructed vertical fields for the axisymmetric disks and spherical electrodes yield comparable *baseline* MSE values of ≤ 7.20E-3. Unlike linear inversion, the inverted Abel solution is restricted to a smaller family of profiles via the constraint of axisymmetry, thereby improving its overall uniqueness and thus accuracy.

For the horizontal field component, overall reconstruction performance is noticeably poorer compared to its vertical counterpart. One possible explanation for this is the comparatively lower signal-to-noise ratios (SNR) for these cases (due to the weaker fields). In general, however, the trends observed across the three methods (for this polarization) remain largely similar, for reasons consistent with those discussed earlier above.

In particular, it is noted that the Abel inversion and machine learning (PC-FDON) methods produce better performance across the 6 different profile types, with MSEs < 2.70E-2 for almost all cases, compared with values of > 4E-2 for linear inversion as shown in Figure 6(b). Abel inversion produces the best performance, in most part due to the stronger signals compared to the other two methods. This advantage is due to the intrinsic nature of the scanning procedure (orthogonal to the direction of laser propagation) for this method (also see equation (3)). From equation (3), the electric field at a given radial location $r$ is the same for $-z$ and $+z$, as $\sqrt{x^2 + z^2}$ is equal to $\sqrt{x^2 + (-z)^2}$. As such, at a given location $x$, the horizontal electric field component that the beam sees will be of a single polarity, as opposed to bipolar for the machine learning and linear inversion cases. And this single polarity often means that these (externally applied) electric fields do not cancel each other out along the beam path, thereby resulting in a stronger signal for the overall profile.

### 3.3. Sensitivity Analysis

#### 3.3.1. Sensitivity to Sparse Data

Sparse data is obtained by down sampling the baseline experimental data, representative of cases where the data is collected at fewer sampling points. In an experiment, this could occur for a variety of reasons, for instance, due to insufficient SNR, or simply to minimize acquisition effort. This section therefore examines the ability of each method to reconstruct the vertical and horizontal components of the electric field from a sparse(r) dataset. A total of 7 different down sampling procedures are tested (see Table 4), wherein methods A to C represent data collected with larger steps; methods D to F represent data with irregular spacing; and method G is used to confirm the importance of the peak location in a few specific horizontal polarization cases.

**Table 4**. List and description of methods used for down sampling the experimental data.

| Method | Down sampling Method |
|---|---|
| Baseline | See number of points in Table 1 |
| A | Sample every 2 points |
| B | Sample every 3 points |
| C | Sample every 5 points |
| D | Retain all points within +/- 2 mm of focus and sample every 3 points |
| E | Retain all points within +/- 5 mm of focus and sample every 3 points |
| F | Retain all points within +/- 10 mm of focus and sample every 3 points |
| G | Retain all points within +/- 10 mm of peaks and sample every 3 points |

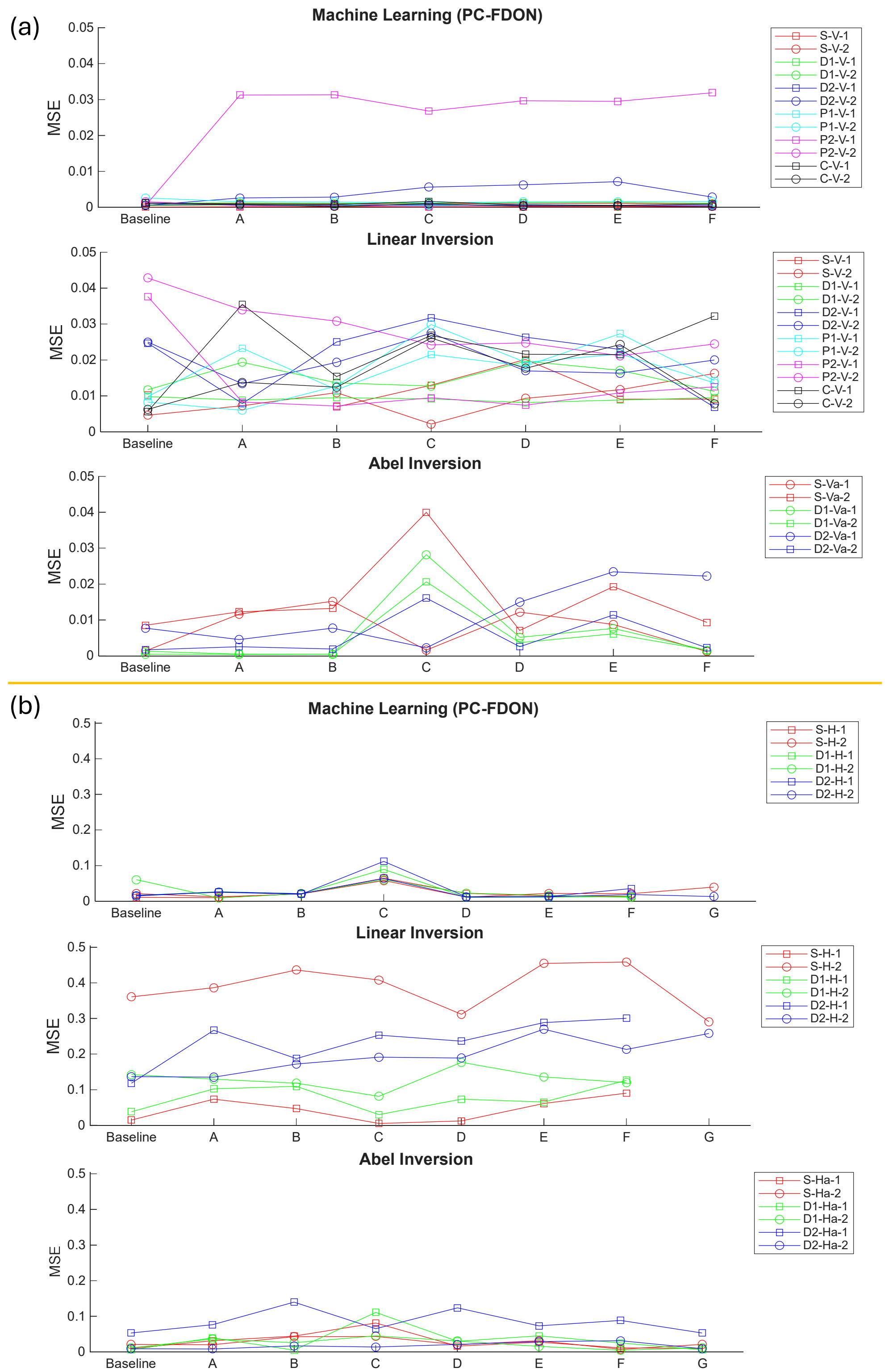

**Figure 6**. MSEs for the reconstruction of the (a) vertical and (b) horizontal components of the electric field corresponding to the six down sampling methods for the vertical polarization (A to F) and seven (A to G) for the horizontal polarization. A total of 6 electrode geometries are examined, the details of which are listed in Tables 1 and 3.

The MSE between the reconstructed components of the electric field and its corresponding simulation results (Laplacian field) for each method is shown in Figure 6. It should be noted that the smoothing parameters are re-optimized for both the linear inversion and Abel inversion methods on every down sampled case (see Appendix B for details). This not only maintains a realistic picture of what would happen in practice but is also found essential to improve the prediction accuracy (compared to the case where these parameters were left unchanged from the original dataset, i.e., *baseline*).

From Figure 6, for the machine learning inversion, it is evident that the MSE for the sparse data reconstruction generally remains comparable with their *baseline* MSE for virtually all six down sampling schemes for most profiles. This can be attributed to the PC-FDON model being inherently feature-driven, i.e., the algorithm has been trained to recognize defining characteristics such as peaks and their locations. Greater weightage is assigned to the profile shape near these features, while the data near the wings assumes a decaying importance. Down sampling therefore has minimal influence on the resulting accuracy as long as these key features are preserved. However, if such vital information is lost, for instance through excessive down sampling, or if the data is extremely noisy, then the model's performance expectedly suffers. A good example can be seen in the 30 x 50 mm electrode with the smaller Rayleigh range (P2-V-1 in Figure 4) as well as the horizontal dataset (Figure 4(b)), where the down sampled dataset results in a loss of essential information near the profile peak, hence leading to poorer accuracy, and thus, elevated MSE (see Figure 6(b)). To be fully transparent, it should be mentioned that the PC-FDON model requires that input profiles be matched to a certain predetermined grid (i.e., a fixed number of 109 points), and where necessary, this is achieved through interpolation. However, this is judged to be more of a procedural requirement and is not expected to bias any of the physical outcomes drawn here.

On the other hand, the larger variations in the MSE across the various down sampling schemes for the linear and Abel inversion methods are because these methods are point-driven. A point-driven method implies that each measurement point is treated as an independent constraint, and every point removed by down sampling (in theory) directly weakens the constraints on the converged solution. In particular, the reconstruction accuracy is generally maintained if features, such as the peak location and profile width are retained. On the other hand, it tends to suffer if these features are omitted, smeared out, or modified due to down sampling. A good example of the former is seen in the vertical component cases, where peaks are located near the beam focus. On the other hand, higher MSEs are recorded for the vertical component with extreme down sampling (i.e. method C; see Figure 4(a)) as well as horizontal component of the Abel inversion (see Figure 4(b)). The down sampling could have altered or removed the features, such as profile width in the case of the vertical component, or the peaks (where they lie away from the beam focus) in the case of the horizontal component. For the horizontal component, down sampling with respect to these peak locations rather than the beam focus (i.e., according to method G) shows comparable MSEs with the *baseline.*

**Table 5**. Parameters of the Lorentzian profiles as well as the true and tested values of the Rayleigh range used in the sensitivity analysis. The definition of the Lorentzian width is identical to that in Table 2. Profile # references the Lorentzian width, $a$ (as defined in equation (2)), used to generate the pre-defined electric fields, while Sub-Profile # serves to identify the true value of the Rayleigh Range.

| Profile # | Sub-Profile # | $a$ [m] | True Rayleigh Range [m] | Tested Rayleigh Range [m] |
|---|---|---|---|---|
| R-1 | A | 0.02 | 0.00146 | 0.00146 to 0.01146; $\Delta z_R$ 0.001 |
| | B | | 0.00646 | |
| | C | | 0.01146 | |
| R-2 | A | 0.05 | 0.00146 | |
| | B | | 0.00646 | |
| | C | | 0.01146 | |
| R-3 | A | 0.1 | 0.00146 | |
| | B | | 0.00646 | |
| | C | | 0.01146 | |

Overall, one may conclude that the reconstruction accuracy is preserved provided that key features are adequately retained. The linear and Abel inversion methods infer this through principles rooted more explicitly in the underlying physics, whereas the machine-learning approach may be viewed as having implicitly *learned* these physical relationships through a carefully structured training procedure.

### 3.3.2. Sensitivity to Rayleigh Range

As mentioned in section 2.2, the Rayleigh range is experimentally measured and input to the various inversion models, given its importance within equation (1). To further understand this relationship, reference E-FISH profiles are synthetically generated based on a pre-defined electric field distribution and laser Rayleigh range.

As shown in Table 5, the pre-defined electric field distributions are generated using three Lorentzian profiles (as defined in equation (2)) of varying FWHM ($a$ in equation (2)). While it is acknowledged that these profiles do not represent the entire family of possible profiles, previous work [12] suggests that these profiles are reasonably similar to vertical field distributions corresponding to canonical electrode geometries, and therefore this investigation is expected to provide relevant insight.

For each Lorentzian field profile, corresponding E-FISH profiles are generated using three different values of the Rayleigh range (designated as the true values) as specified under True Rayleigh Range in Table 5. Each E-FISH profile is subsequently inverted using the three reconstruction methods, with the assumed Rayleigh range systematically varied over a specified ‘test’ range (indicated in the rightmost column of Table 5) to assess its influence on reconstruction accuracy. The idea is to mimic a possible scenario whereby an incorrect estimate of the Rayleigh range is used for the reconstruction. This is further supported by sensitivity analyses for the linear inversion and Abel inversion methods (see Appendices C and D, respectively).

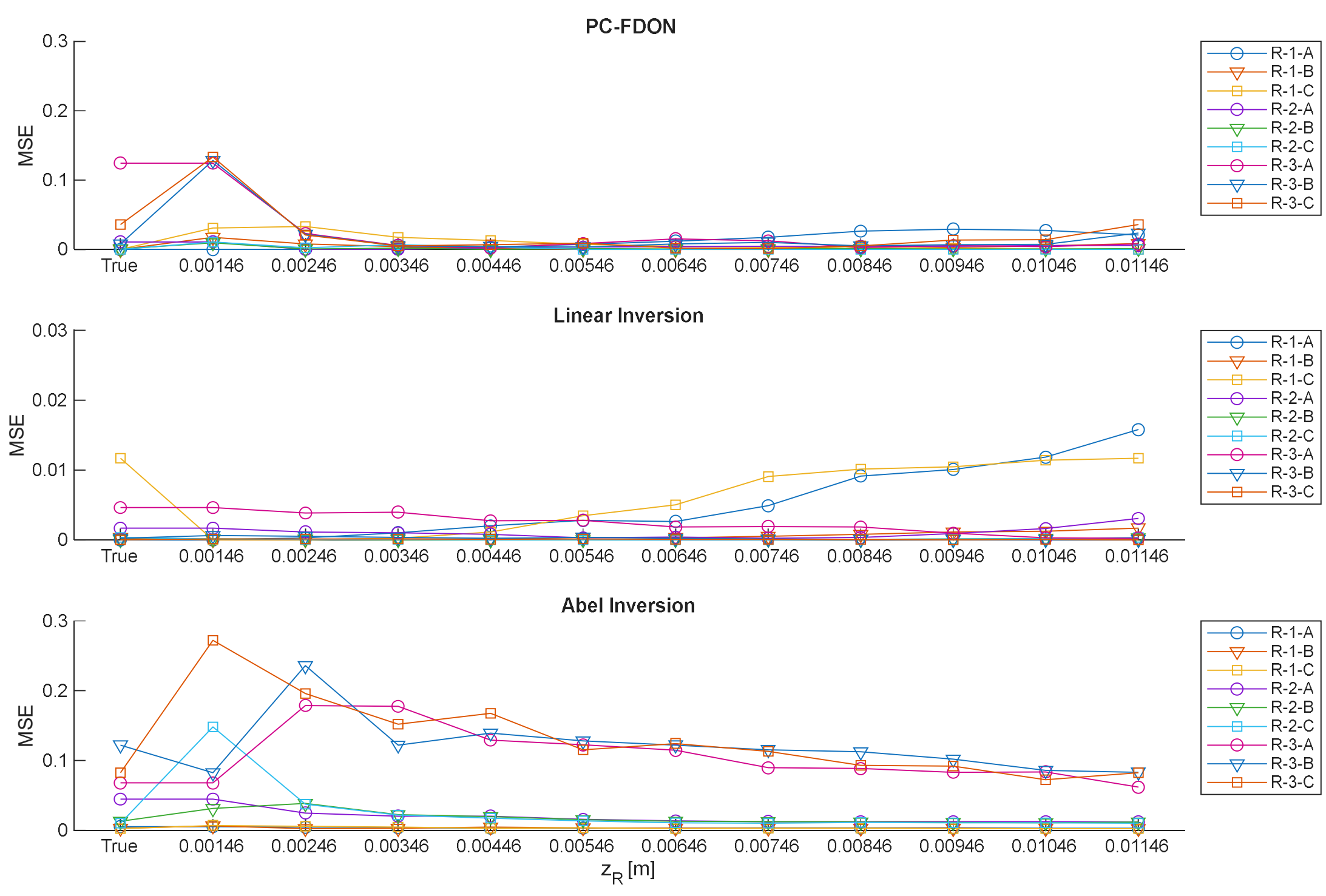

**Figure 7**. MSEs for the reconstruction of the vertical component of the electric fields with varying Rayleigh range: (top) for machine learning; (center) for linear inversion; and (bottom) for Abel inversion model. The abscissa denotes the tested values of the Rayleigh range for each Lorentzian profile. True refers to the MSE for the electric field reconstructed using the same Rayleigh range as that used to generate the E-FISH signal. The legend refers to the Profile # followed by the Sub-Profile # as listed in Table 5.

To quantify this (Rayleigh range) sensitivity, the MSE between each reference Lorentzian profile and the reconstructed electric field profile resulting from the varied Rayleigh range is compared in Figure 7 for all three reconstruction methods. For the linear and Abel inversion methods, recovery is performed with the same smoothing parameters applied and the same Lorentzian profile width is used as the initial guess.

Following Figure 7, it is evident that both the machine learning method and linear inversion are sensitive to the Rayleigh range. The PC-FDON model requires the Rayleigh range as an input to the model and has been trained specifically to account for changes in this parameter. As seen in Figure 4 (see for example P1-V-1 and P1-V-2), applying a different Rayleigh range for the same electric field profile can result in an entirely distinct E-FISH distribution. In addition, the Rayleigh range is also used to non-dimensionalise the spatial coordinate of the input and output profiles. Considering that the model operates over a fixed input domain (i.e., $|z'| \leq 50$ (where $z' = z/z_R$)), large errors in the Rayleigh range may result in increased reconstruction errors as key features fall out of this prescribed domain, such as R-3-C for the machine learning model).

Likewise, for the linear inversion model, as the reconstruction is based on minimising the MSE between the E-FISH signal and forward integration using equation (1), an incorrect Rayleigh range may also result in an incorrect forward integration of the electric field. More importantly, even if the residual between the E-FISH profiles is minimised, the resulting reconstructed electric field may be vastly different from the true electric field, due to an incorrect Rayleigh range, similar to the non-uniqueness arguments presented in section 3.2.

A theoretical basis for understanding the sensitivity of the E-FISH signal to the change in Rayleigh range is provided using the equation derived in Appendix C. This analysis holds for both the machine learning and the linear inversion methods, as the equation is based on equation (1) which the signals for both methods are derived from.

Similar to the profiles in Table 5, the three different Lorentzian widths (corresponding to R-1, R-2, and R-3) are also used to generate profiles for $E_{ext}(z)$. The sensitivity values are then calculated following equation (C.2), normalised by the sensitivity at $z = 0$. These normalised sensitivities at various $z$-locations and for various Rayleigh range values are shown in Figure 8(a).

It is evident from Figure 8(a) that the signal at various locations is affected differently with change in the Rayleigh range, although the effect is symmetrical with respect to the focus. While the signal at the focus (0 m) may vary negligibly with change in Rayleigh range due to the low sensitivity, the signal at other locations varies considerably across the Rayleigh range being considered. As such, a significant variation in signal with Rayleigh range change is therefore expected, as also earlier observed in Figure 7.

On the other hand, for the Abel inversion method, the reconstructed field is also sensitive to changes in the Rayleigh range, albeit for a slightly different reason. As shown in equations (3) and (4), the Rayleigh range appears in the matrix $W$, which contains the constants relating the reconstructed electric field to the E-FISH signal. Errors in the assumed Rayleigh range therefore introduce errors into $W$, potentially distorting the reconstructed electric field and its attendant features, compared to the reference (i.e., Laplacian) electric field.

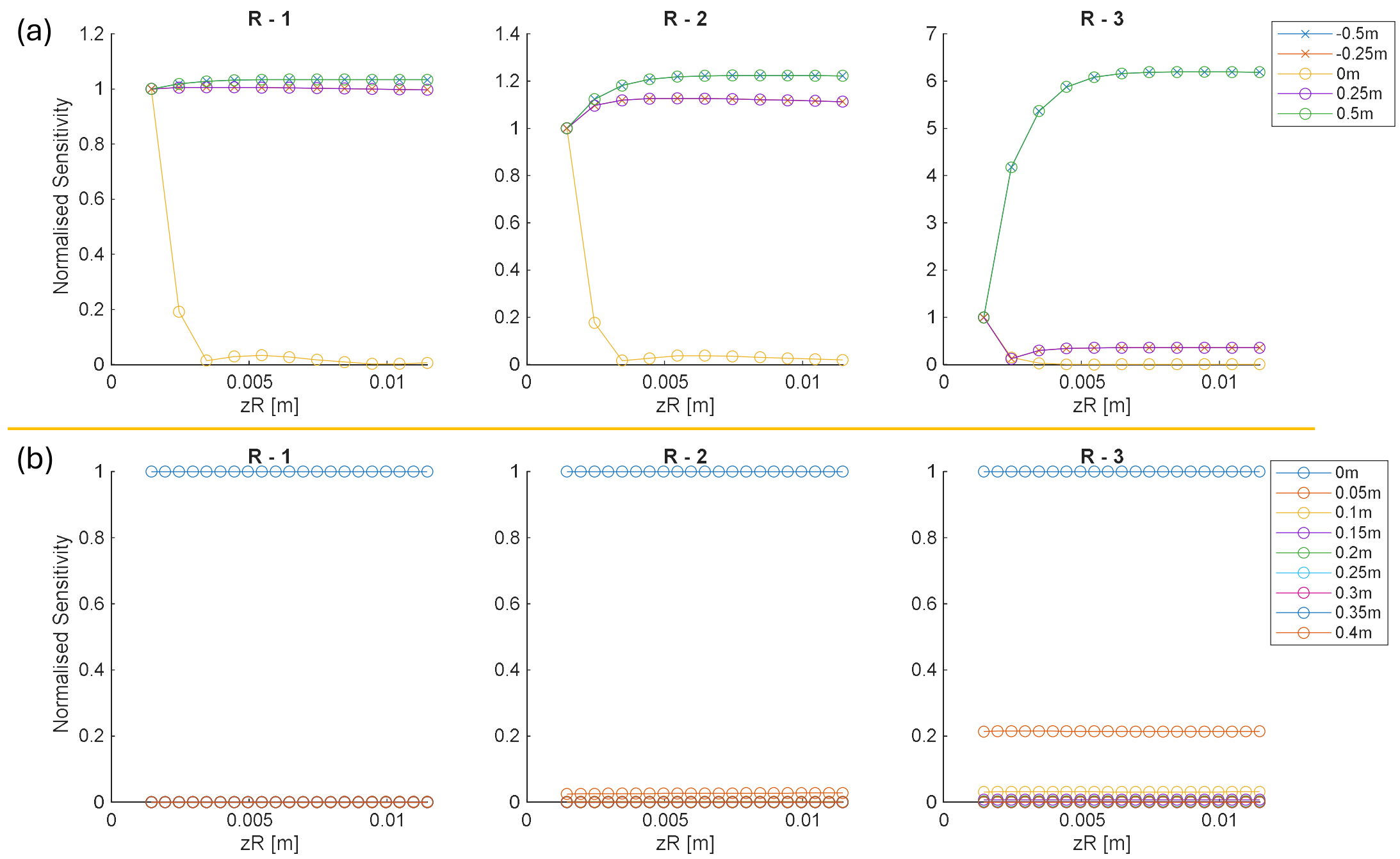

**Figure 8**. Normalised sensitivity values for 3 different Lorentzian profiles and for various Rayleigh range values, for (a) machine learning and linear inversion methods and (b) Abel inversion. For plot (a), the legend indicates the shift of the profile peak from the focus (0 m) and for plot (b), the legend shows the various radial locations.

To confirm this observation, the sensitivity of the E-FISH signal to the change in Rayleigh range can be theoretically analysed, following equation (D.7). Three different $E_{ext}\left(\sqrt{x^2+z^2}\right)$ profiles are also generated corresponding to profiles R-1, R-2, and R-3 in Table 5, and thereafter, are substituted to equation (D.7). After normalising the values with the sensitivity at $x$ = 0, these values are shown in Figure 8(b) for various $x$-locations and Rayleigh ranges.

From Figure 8(b), it is evident that with a thinner profile, such as in R-1, the signals are affected almost equally at various $x$-locations. This results in the method being insensitive to the change in Rayleigh range. However, as the profiles become wider, the signals at the various $x$-locations are affected in differing values (see R-2 and R-3 in Figure 8(b)), such that the method will exhibit higher sensitivity to the change in Rayleigh range.

### 3.3.3. Sensitivity to Choice of Initial Profile

Unlike the other two methods, the PC-FDON model does not require the selection of an initial profile to initiate the solution process, and the output solution is also invariant for a given input. This section therefore examines the effect of the (choice of) initial profile on the solution quality for both linear and Abel inversion.

The choice of these initial profiles is guided by the known Laplacian electric field distribution, and the reconstructed electric field is compared as this initial guess is varied. This procedure is performed for a given smoothing parameter, Rayleigh range, and E-FISH signal.

From Figure 9, it is evident that the chosen profile affects only the reconstructed electric field (*Opt.*) produced by the linear inversion model. On the other hand, Figure 9(b) shows that the Abel inversion method is insensitive to this initial choice. The first plot with "Inf" guess width indicates a constant value (this can be for instance, all zeros or ones) supplied as the initial profile, and compared to the other two guess widths, results in an almost identical reconstructed electric field.

The differing sensitivity to initial profile comes from the different optimization and/or iteration process for the two methods. The linear inversion iterates by comparing the forward E-FISH of the (input) electric field with the experimental data until their residual becomes acceptable, thereafter designating *this* electric field profile as the reconstructed solution. Therefore, an appropriate initial guess is important as the problem is less constrained (due to the issue of non-uniqueness discussed in section 3.2) and may admit non-physical or inaccurate solutions. On the other hand, the Abel inversion solves equation (4) iteratively to minimize the error between its two sides. The use of absolute values introduces non-linearity into the equation, thereby necessitating an initial guess for the iterative solution. As illustrated in Figure 9(b), however, its dependence on the choice of initial profile is significantly weaker compared to the linear inversion method. This is because each radial measurement depends only on the field at larger radii, leading to an essentially unique solution.

As such, selecting an appropriate initial profile for the given electric field is of decisive importance, in particular for the linear inversion model. This could pose a problem, in cases such as a plasma, where the electric field profiles of interest are completely unknown. In such instances, an approach following that in section 3.5 is suggested.

#### 3.3.4. Sensitivity to Noise

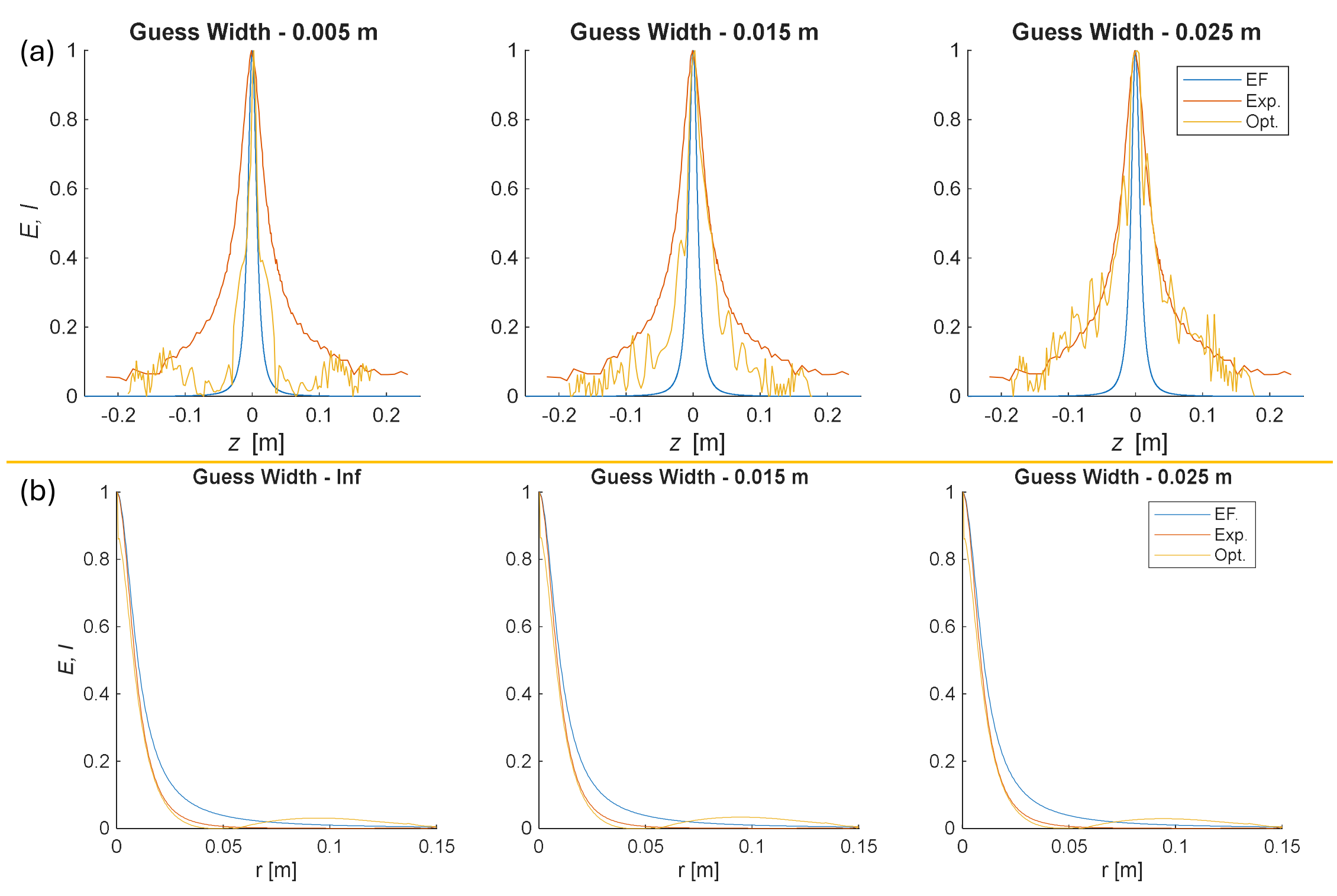

**Figure 9**. Sensitivity to initial profile for (a) linear inversion method and (b) Abel inversion method. The different choices of initial profiles are indicated in each of the plot titles. EF refers to the Laplacian electric field; Exp. refers to the experimental E-FISH signal; Opt. refers to the reconstructed electric field using either the linear inversion or Abel inversion method.

Although experimental data inherently contains noise, a systematic analysis of each method's ability to accurately reconstruct the electric field profiles for noisy data is necessary. To do so, the normalised, numerically integrated E-FISH signal, generated from a Lorentzian profile, is modified with three different Gaussian noise magnitudes. Details of the Lorentzian profile and the Gaussian noise magnitudes are listed in Table 6.

From Figure 10, it appears that the linear inversion method is more sensitive to noise, as compared to the machine learning and Abel inversion models. The added noise results in an MSE that is up to 100 times higher with linear inversion, while the other two methods generally exhibit MSEs that remain closer to their baselines. For the machine learning model, this resilience may be attributed to two main reasons. As discussed in section 3.3.1, since the algorithm is trained to recognize key features, reconstruction accuracy is maintained if such features are retained. The other explanation is directly tied to the fact that the model has been trained (or preconditioned) on noisy data. As for the Abel inversion method, we surmise that this consistent performance is due to the axisymmetric constraint which acts to restrict the solution space that noise can bias the output towards.

**Table 6**. Parameter for the Lorentzian profile, $a$ (as defined as in equation (2)), added Gaussian noise magnitude, and the corresponding MSE for the reconstructed electric field. *MSE M, MSE L* and *MSE A* refer to the MSEs for the reconstruction done using the machine learning, linear inversion, and Abel inversion method respectively.

| # | $a$ [m] | Noise Magnitude [%] | MSE M | MSE L | MSE A |
|---|---|---|---|---|---|
| N-B | 0.02 | 0 | 1.96E-4 | 6.25E-4 | 6.40E-5 |
| N-1 | | 5 | 1.60E-3 | 6.05E-2 | 2.89E-4 |
| N-2 | | 10 | 3.72E-2 | 7.84E-2 | 1.76E-3 |
| N-3 | | 15 | 3.97E-3 | 6.05E-2 | 2.81E-3 |

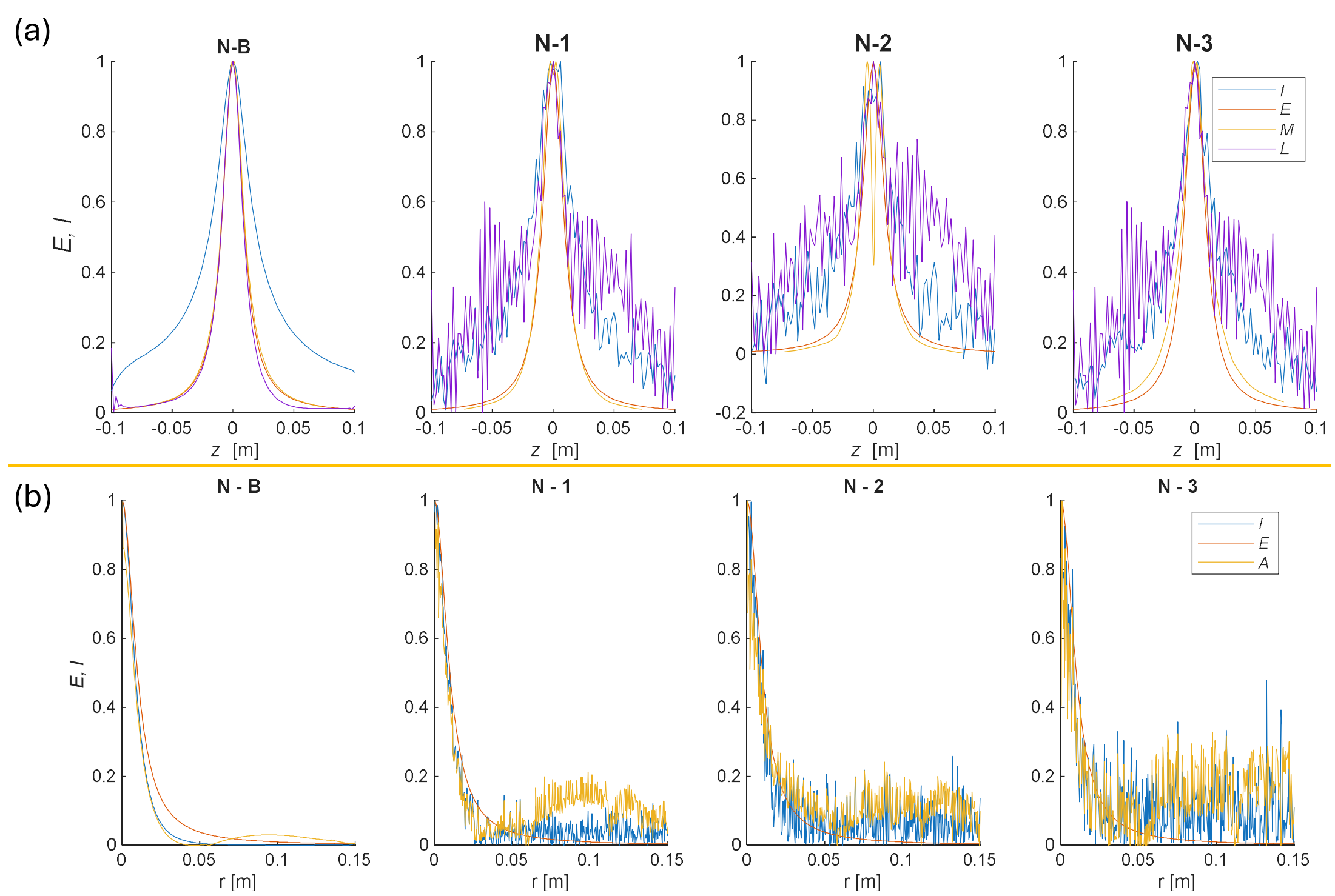

**Figure 10**. Sensitivity to noise, for a given profile and smoothing parameter for (a) machine learning and linear inversion methods; and (b) Abel inversion method. On plots (a) and (b), *I* refers to the numerically integrated E-FISH signal from the electric-field *E*. For plot (a), *M* and *L* are the reconstructed electric fields using the machine learning (PC-FDON) and linear inversion model respectively; and for plot (b), *A* is the reconstructed electric field using the Abel inversion method. The title on each plot refers to the # on Table 6.

The above notwithstanding, it should be cautioned that this superior performance may deteriorate if key features are obscured, shifted, and or modified by excessive noise (e.g. plot N-2 of Figure 10(a)), or, in the case of machine learning, if this noise character differs substantially from that utilized during its training. Maintaining good accuracy therefore demands a certain judgment on the acceptable level of data quality during the collection process.

### 3.4. Asymmetric Profile Shapes

Although the experimental data obtained from scanning the horizontal and vertical components of the electric field profile provide a wide range of profile widths and shapes, asymmetric profiles are intentionally absent from this comparison. Asymmetric profiles refer to cases where the electric field profiles are not symmetric about the beam focus (i.e. $z = 0$). Since the Abel inversion method presupposes axisymmetry, only the machine learning model and the linear inversion method can be compared.

Three arbitrary asymmetric field profiles, as listed in Table 7, are generated by summing two Lorentzian profiles (*Lorentzian A* and *Lorentzian B*, following equation (2)), each with its own width ($a$) and peak shift ($c$) (see $E$ in Figure 11). The resulting profiles are numerically integrated according to equation (1) to obtain their corresponding E-FISH signals (see $I$ in Figure 11), which are in turn input to the machine learning and linear inversion models, using the parameters in section 2.4. The reconstruction results are shown as *M* for machine learning and *L* for linear inversion in Figure 11.

**Table 7**. Parameters for the arbitrarily chosen asymmetric profiles ($a$ and $c$ are as defined in equation (2)) and the corresponding MSE for the reconstructed electric field. *MSE M* and *MSE L* refer to the reconstruction done using the machine learning and linear inversion methods respectively.

| # | Lorentzian A | | Lorentzian B | | MSE M | MSE L |
|---|---|---|---|---|---|---|
| | $a$ [m] | $c$ [m] | $a$ [m] | $c$ [m] | | |
| 1 | 0.01 | -0.006 | 0.006 | 0.008 | 5.28E-3 | 2.14E-2 |
| 2 | 0.006 | -0.008 | 0.004 | 0.002 | 8.32E-2 | 3.13E-4 |
| 3 | 0.012 | -0.004 | 0.004 | 0.012 | 5.51E-2 | 6.40E-4 |

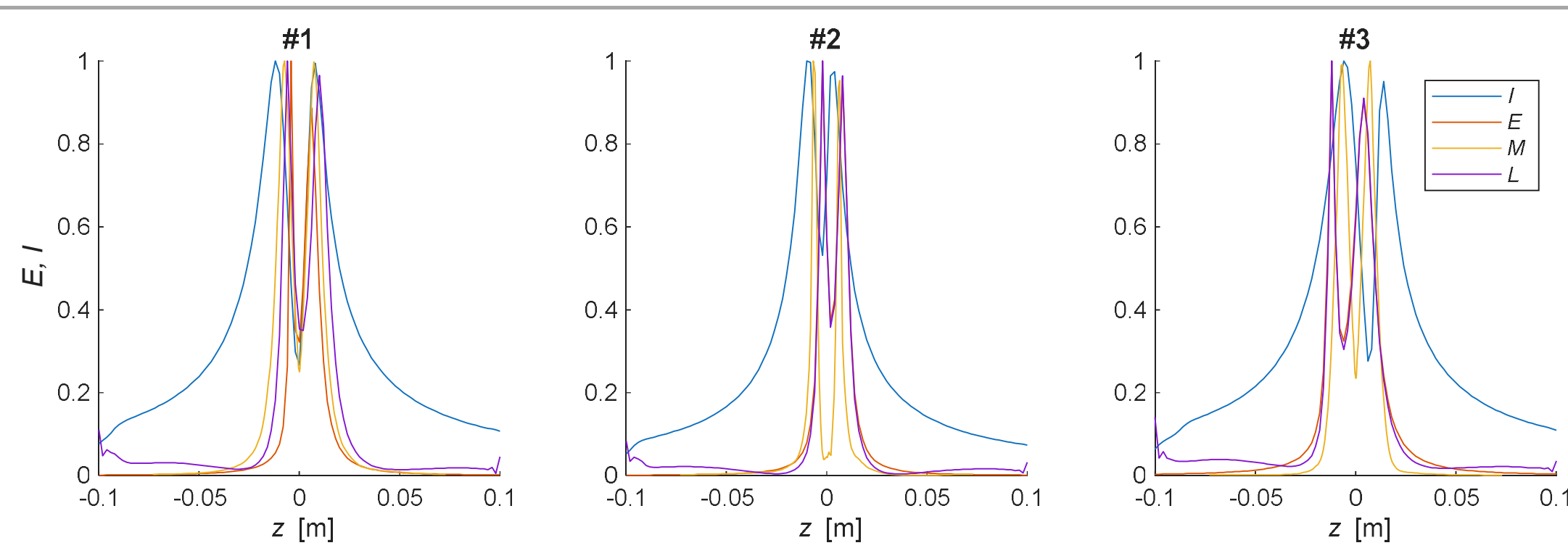

**Figure 11**. Reconstruction results for the three (#1-3) asymmetric electric fields from Table 7. $I$ refers to the numerically integrated E-FISH signal from the electric-field $E$; $M$ and $L$ are the reconstructed electric fields using the machine learning (PC-FDON) and linear inversion models respectively.

Comparing the MSE for the respective methods in Table 7 and the observed fit in Figure 11, it is noted that linear inversion performs better overall in asymmetric profile reconstruction, at least for the Lorentzian distributions considered here. This is due to an inherent limitation of the machine learning model, whose training dataset has been limited to symmetric distributions. In these cases, a proposed solution is to use the machine learning model output as the initial guess (or input) for the linear inversion model to further optimize the accuracy of the reconstructed electric field (see also section 3.5).

### 3.5. Reconstruction of Electric Field Profiles of Unknown Shape

The foregoing sections establish that the linear inversion method is sensitive to the initial input profile, which clearly poses a problem when applied to real plasmas, and it is therefore important to find suitable reconstruction approaches for such cases. This broadly extends to profiles that are initially unknown, or even asymmetric profiles that are typically beyond the capability of the PC-FDON model.

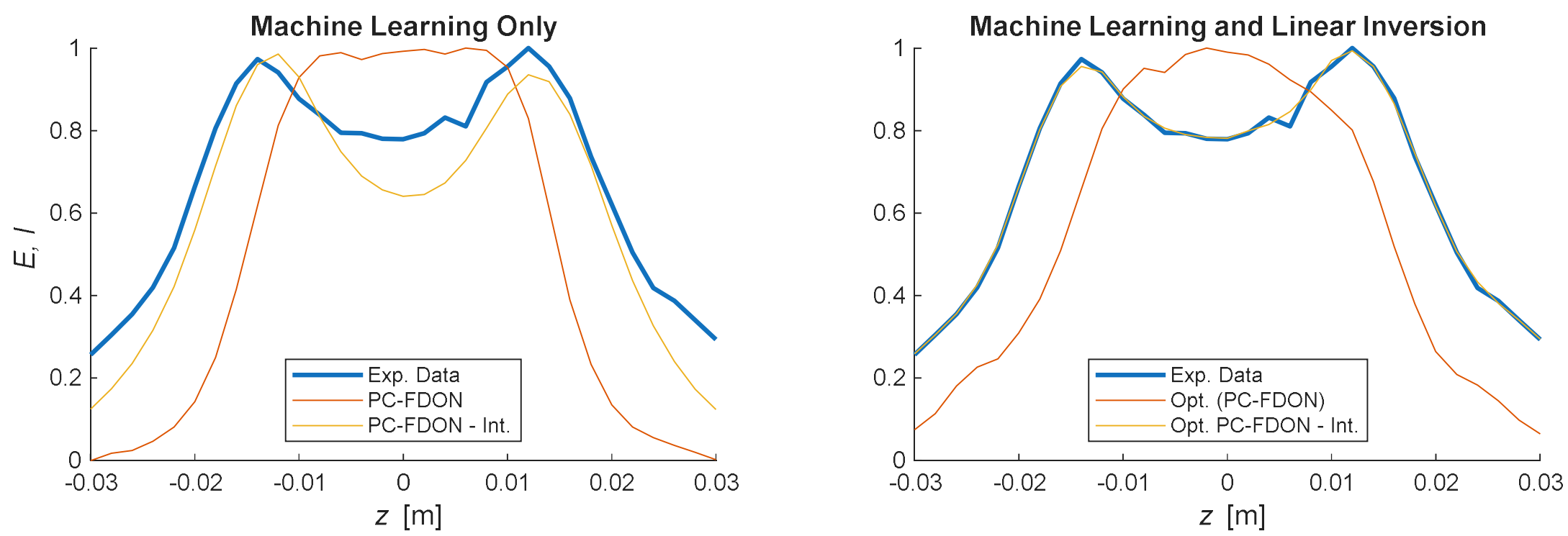

**Figure 12**. Reconstructed electric field using only the machine learning model (PC-FDON) (left) and after using the PC-FDON output as the initial guess for the linear inversion model (right). Exp. Data refers to the experimental E-FISH signal; *PC-FDON* refers to the reconstructed electric field using the PC-FDON model; *PC-FDON – Int.* refers to the numerically integrated E-FISH signal from *PC-FDON*; *Opt. (PC-FDON)* refers to the reconstructed electric field using the linear inversion method with PC-FDON as the initial guess; *Opt. PC-FDON – Int.* refers to the numerically integrated E-FISH signal from *Opt. (PC-FDON)*; *E* refers to the electric field; *I* refers to the E-FISH signal intensity.

Using the E-FISH signal obtained from the electric field generated by the dielectric barrier discharge as an example (see Figure 12), the experimental E-FISH signal (*Exp. Data*) is reconstructed using the machine learning method as a first pass. This output (*PC-FDON*) is subsequently used as the initial profile that is passed to the linear inversion method. Such an approach is also similar to that used in [18, 19]. The MSE between the experimental E-FISH signal and the E-FISH signal obtained by numerically integrating the *PC-FDON* (*PCFDON – Int.*) is 1.74E-2. This MSE is significantly higher than the value of 1.08E-4 from the field reconstructed by the linear inversion method based on the *PC-FDON* output as the initial guess (*Opt. PCFDON – Int.*). The use of the linear inversion method, by leveraging the machine learning solution, indeed decreases the error. Nevertheless, it is noted that the reconstructed electric field is non-unique, and thus, further validation of the electric field spatial distribution through other means may be necessary.

### 3.6. Appropriate Method Selection for Electric Field Reconstruction

This final section discusses 4 implementation aspects, namely parameter training and selection; reconstruction speed; required grid density; and signal strength. The key findings are summarised in

Table 8. This collectively serves as a practical aid in selecting the appropriate inversion method for a given experimental setup.

### 3.6.1. Parameter Training and Selection

Prior to using the machine learning model to reconstruct electric fields, it must be trained with a dataset. This dataset must include the types of profile and parameters, for instance Rayleigh range and

**Table 8**. Comparison of three methods based on four metrics.

<table>
<tr><th>Metric</th><th>Machine Learning (PC-FDON)</th><th>Linear Inversion</th><th>Abel Inversion</th></tr>
<tr><td rowspan="2">Parameter Training and Selection</td><td rowspan="2">Requires training</td><td colspan="2">• Does not require training<br>• Requires optimization of smoothing parameters</td></tr>
<tr><td>Requires appropriate initial profile</td><td>Not sensitive to initial profile</td></tr>
<tr><td>Reconstruction Speed</td><td>Within seconds</td><td colspan="2">Depending on grid density</td></tr>
<tr><td>Required Grid Density</td><td>• Fixed input grid<br>• May pose issues for noisy data</td><td>Dense grids preferred</td><td>• Only for axisymmetric fields<br>• Collect only half domain</td></tr>
<tr><td>Signal Intensity</td><td colspan="2">Weaker signal for horizontal polarization</td><td>Stronger signal for horizontal polarization</td></tr>
</table>

wave vector mismatch, expected for the electric fields that are to be reconstructed. The training may take up to 1.8 hours using an NVIDIA H100 GPU. Once deployed (i.e., trained, tested and validated), however, the reconstruction process completes within seconds for each dataset.

On the other hand, the linear inversion and Abel inversion models do not require pre-training. However, for the linear inversion, the smoothing parameters must be optimised, which may take up to 20 hours on a 12-core CPU with a clock speed of 2.20 GHz, depending on the grid size and parameter range (see also section 3.6.2). As for the Abel inversion method, the smoothing parameters must also be optimised (depending on the grid size and range of parameters), but the process usually takes no more than 20 minutes (on a 6-core CPU with the same clock speed of 2.20 GHz) for each case. If the once-off training cost is neglected, the PC-FDON machine learning model requires the shortest time to generate a solution, followed by Abel inversion and linear inversion.

Additionally, the use of the linear inversion method requires an initial profile, which must be of comparable width (or shape) to the expected electric field (see section 3.3.3). In the case of unknown electric fields, other methods may be leveraged to obtain the initial guess (see section 3.5). With the given parameters, the time taken for the reconstruction process varies according to the grid density.

### 3.6.2. Required Grid Density

The PC-FDON model has a fixed input grid, but as earlier discussed in section 3.2, its performance is fundamentally dictated by whether sufficient information on the peak shapes and locations is retained, rather than the number of acquired data (or grid) points. (Recall that, regardless of the acquired grid density, the data are interpolated onto the fixed 109-point input grid required by the model.) Although this allows the model to remain relatively insensitive to variations in grid density up to a certain extent, issues arise when the input data loses information on the peak shape (see section 3.3.1) after interpolation or when the data are subject to substantial noise (see section 3.3.4).

For the linear inversion model, denser grids are generally preferred, especially for noisy data as well as profiles with sharp spatial gradients. As a trade-off, however, the time taken for the inversion process may be longer (see also section 3.6.1) as this scales directly with the input grid density.

Finally, for the Abel inversion model, only signals along half of the spatial domain are required, as compared to the linear inversion model. This reduced data collection requirement hinges directly on the axisymmetric assumption imposed for this inversion method. Care must therefore be taken to ensure this important condition is satisfied. Simple checks could include confirming that the presence of electrode holders (or experimental components) does not bias the symmetry of the underlying electric field distribution. As a cautionary note, it should be further emphasized that maintaining axisymmetry of the Laplacian field does not necessarily guarantee that this condition continues to apply in a plasma.

#### 3.6.3. Signal Strength

In general, the signal strength associated with a given electrode geometry for all three methods remains similar for the vertical polarization. For the horizontal component, in the case of machine learning and linear inversion methods, as the E-FISH signal intensity follows equation (1), the positive and negative components of the electric field cancel out, resulting in a weak(er) signal. On the other hand, the intensity for the Abel inversion follows equation (3). In this case, the positive and negative components of the electric field do not cancel, resulting in a signal intensity that is typically up to 5 times stronger.

The Abel inversion method, by virtue of its scanning direction (*orthogonal* to the laser path), often allows regions closer to the electrode surfaces to be interrogated. In contrast, displacing the field *along* the direction of laser propagation may cause the expanding beam to be partially obstructed by other upstream (or downstream) components of the setup.

## 4. Conclusion

Several reconstruction methods have been recently proposed to address the measurement uncertainty associated with the E-FISH diagnostic due to the path-integrated nature of its signal production. These entail considerably distinct implementations but have not been previously assessed against one another. This work therefore systematically compares three such methods: machine learning, linear inversion and Abel inversion. The former makes use of a PC-FDON model developed in our group while the latter two methods are reproduced in-house following their published formulations. The comparison is facilitated by measuring E-FISH signals from four canonical electrode geometries (sphere, disk, plate and cylinder) under electrostatic conditions, and for both the vertical and horizontal components of the electric field.

Using the MSE of the reconstructed solutions relative to the simulated (true) Laplacian field, all three methods are found capable of accurate profile recovery, although no method emerges as the most optimal. Some key insights are as follows:

(i) we regard the PC-FDON model as the most suitable for general use, although its applicability (at present) remains limited to symmetric and antisymmetric profiles. This is related to its machine learning architecture, which broadly restricts the model to such profiles which the algorithm has been trained and validated on.

(ii) Linear inversion is in principle applicable to any profile shape but yielded higher MSEs overall, a consequence of the non-uniqueness of solutions constrained only by agreement with the measured signal. Selection of an appropriate initial profile is also requisite for good reconstruction.

(iii) Abel inversion matches the accuracy of PC-FDON for the vertical field component, and produces the best results for the horizontal component, in part due to the stronger signals for the latter polarization. The method's main drawback is the required assumption of field axisymmetry, but in applications where this condition is satisfied, Abel inversion would certainly be a sensible option.

When tested on E-FISH profiles with fewer datapoints (i.e., sparse datasets) and added noise, reconstruction accuracy is observed to be directly related to how well the key features (e.g. profile peaks) are preserved. The machine learning and Abel inversion methods generally outperform linear inversion. The PC-FDON appears capable of recognizing these important features through its underlying convolutional neural networks, while the inverse Abel method performs better due to the smaller solution space that the condition of axisymmetry enforces.

We note that both Abel and linear inversion operate on a largely similar mathematical approach and optimization objective, except that one may be viewed as a more constrained version of the other. The requirement of axisymmetry limits the applicability of Abel inversion but provides improved reconstruction performance in return.

For asymmetric profiles which lie beyond the capability of the PC-FDON model, linear inversion performed significantly better. Given this complementary relationship, we propose a combined approach whereby the machine learning output serves as the initial guess for linear inversion. Applied to an SDBD with an a priori unknown field profile, this two-step approach reduces the MSE between the measured and forward-calculated E-FISH signals by about two orders of magnitude relative to machine learning alone.

In terms of implementation, machine learning incurs a one-time training cost, but reconstruction is essentially immediate (approximately a few seconds) and demands little additional computational effort from the user. Abel and linear inversion require no training, but optimization of their regularization parameters and initial profile selection must be performed for each dataset, costing up to one and three orders of magnitude more respectively than machine learning. Taken together, the assessment of reconstruction accuracy and robustness, sensitivity to the various input parameters, and practical aspects of implementation provides a firm basis for selecting an appropriate reconstruction method for a given experimental configuration and field profile.

In summary, while PC-FDON offers the best overall balance for general-purpose reconstruction, linear and Abel inversion retain important advantages where greater flexibility in profile shape or axisymmetry, respectively, can be exploited.

## Acknowledgements

We wish to acknowledge the support from a Singapore Ministry of Education (MOE) grant (22-4973-A0001), and a Singapore MOE PhD research scholarship for Mr. Edwin Setiadi Sugeng.

## Appendix A

The chosen first derivative (*Dx*) as well as the second derivative (*Dx2*) regularisation parameters for the linear inversion method are listed in Table A-1, arranged according to each electrode type. The smoothing parameters, *c*, for the Abel inversion method are listed in Table A-2, also arranged according to electrode type. The numbers in the first column (#) refer to the numbering in Table 3.

**Table A-1**. Regularization parameters used in the linear inversion method. *Dx* refers to the first derivative smoothing parameter while *Dx2* refers to the second derivative smoothing parameter. Both *Dx* and *Dx2* are in powers of 10, with zero indicated as Z.

| # | *Dx* | *Dx2* | # | *Dx* | *Dx2* | # | *Dx* | *Dx2* | # | *Dx* | *Dx2* |
|---|---|---|---|---|---|---|---|---|---|---|---|
| S-V-1 | -3.5 | -4 | D2-V-2 | -3 | -4 | C-V-1 | Z | 2 | D1-H-1 | 1 | 1 |
| S-V-2 | -4 | Z | P1-V-1 | 1.5 | 0 | C-V-2 | Z | 2 | D1-H-2 | 1 | 1 |
| D1-V-1 | -3.5 | -2.5 | P1-V-2 | -4 | -2.5 | S-H-1 | -3.5 | Z | D2-H-1 | 1 | 1 |
| D1-V-2 | -2.5 | 0 | P2-V-1 | Z | -4 | S-H-2 | Z | -2.5 | D2-H-2 | 1 | 1 |
| D2-V-1 | -4 | -4 | P2-V-2 | -4 | -1 | | | | | | |

**Table A-2**. Regularization parameters used in the Abel transform method. c refers to the smoothing parameter. The *c* values are factor of $10^{-2}$.

| # | *c* | # | *c* | # | *c* | # | *c* | # | *c* |
|---|---|---|---|---|---|---|---|---|---|
| S-Va-1 | 2.5 | D1-Va-2 | 8.75 | S-Ha-1 | 1.25 | D1-Ha-1 | 1.25 | D2-Ha-1 | 1.25 |
| S-Va-2 | 18.75 | D2-Va-1 | 17.5 | S-Ha-2 | 2.5 | D1-Ha-2 | 2.5 | D2-Ha-2 | 2.5 |
| D1-Va-1 | 6.25 | D2-Va-2 | 8.75 | | | | | | |

## Appendix B

The chosen first derivative (*Dx*) as well as the second derivative (*Dx2*) regularisation parameters applied for the linear inversion method are listed in Table B-1, arranged according to the electrode type and down sampling method. The smoothing parameter, c, for the Abel inversion method is listed in Table B-2, also arranged according to electrode type and down sampling method. For each table, the numbers in the first column (#) refer to the numbering in Table 3.

## Appendix C

**Table B-1**. Regularization parameters used in the linear inversion method. *Dx* refers to the first derivative smoothing parameter, while *Dx2* refers to the second derivative smoothing parameter, and the alphabets A to G refers to the down sampling type in Table 4. *Dx* and *Dx2* are in powers of 10, with zero indicated as Z.

| # | A | | B | | C | | D | | E | | F | | G | |
|---|---|---|---|---|---|---|---|---|---|---|---|---|---|---|
| | *Dx* | *Dx2* | *Dx* | *Dx2* | *Dx* | *Dx2* | *Dx* | *Dx2* | *Dx* | *Dx2* | *Dx* | *Dx2* | *Dx* | *Dx2* |
| S-V-1 | Z | Z | 1.5 | 2 | 1.5 | 2 | 2 | 2 | -3 | Z | -3.5 | -2 | | |
| S-V-2 | Z | Z | Z | Z | Z | Z | -4 | -4 | -3.5 | -2.5 | -2.5 | -3 | | |
| D1-V-1 | 1 | -3.5 | -3.5 | Z | Z | Z | Z | Z | Z | -3.5 | Z | -3.5 | | |
| D1-V-2 | Z | -3.5 | 1.5 | 0.5 | Z | Z | -3 | -3 | -4 | Z | -4 | -4 | | |
| D2-V-1 | Z | -3.5 | -4 | -4 | -4 | Z | Z | Z | -2 | 2 | -3.5 | -3.5 | | |
| D2-V-2 | -3 | -2 | -4 | -1 | -4 | Z | Z | -1 | Z | -4 | -3 | Z | | |
| P1-V-1 | -2.5 | Z | Z | -4 | Z | -2 | -3.5 | -2.5 | -3 | -3 | -2.5 | -3.5 | | |
| P1-V-2 | -4 | Z | -1 | -1 | -3 | -4 | -2 | -2 | Z | -3 | Z | -2.5 | | |
| P2-V-1 | Z | -3 | -3.5 | -2.5 | 0 | 2 | -2 | 2 | Z | 2 | -3.5 | -2.5 | | |
| P2-V-2 | -2 | -2 | -4 | -2.5 | -3 | -2.5 | -4 | -1 | -3.5 | -4 | -3.5 | -2.5 | | |
| C-V-1 | -2 | 2 | -4 | -2.5 | -1 | 2 | Z | 0.5 | -1.5 | 0.5 | -2 | -1 | | |
| C-V-2 | -0.5 | 1.5 | -2.5 | -1.5 | 0.5 | 2 | Z | 2 | Z | 1 | 0 | 2 | | |
| S-H-1 | -1 | -3 | 0 | 1.5 | -1 | -2.5 | -3 | -4 | 1 | -1 | 0 | -1 | | |
| S-H-2 | -3 | 0.5 | -4 | -3.5 | -2.5 | 1.5 | -2.5 | 2 | -1 | 1.5 | -2 | 0.5 | -4 | -2 |
| D1-H-1 | Z | Z | 2 | -1.5 | Z | Z | Z | -3.5 | Z | Z | -2 | -3.5 | | |
| D1-H-2 | Z | -1 | 1.5 | 1 | 0 | 0 | 0.5 | -2.5 | 1 | -2.5 | 0 | 2 | | |
| D2-H-1 | -3.5 | -2.5 | -2 | 2 | -0.5 | -2 | -1.5 | -2.5 | -2 | -1.5 | -3.5 | Z | | |
| D2-H-2 | -2 | -0.5 | -2.5 | -3 | 2 | 2 | 2 | -0.5 | -1.5 | 2 | 1.5 | 2 | -2 | -4 |

**Table B-2**. Smoothing parameters, *c*, used in the Abel transform method. The alphabets A to G refer to the down sampling scheme in Table 4. The values are factor of $10^{-2}$.

| # | A | B | C | D | E | F | G |
|---|---|---|---|---|---|---|---|
| S-Va-1 | 1.25 | 1.25 | 25 | 1.25 | 1.25 | 1.25 | |
| S-Va-2 | 1.25 | 0 | 1.25 | 1.25 | 5 | 0 | |
| D1-Va-1 | 5 | 3.75 | 6.25 | 1.25 | 0 | 2.50 | |
| D1-Va-2 | 0 | 0 | 0 | 1.25 | 1.25 | 1.25 | |
| D2-Va-1 | 1.25 | 0 | 1.25 | 1.25 | 1.25 | 0 | |
| D2-Va-2 | 1.25 | 3.75 | 3.75 | 1.25 | 0 | 2.50 | |
| S-Ha-1 | 1.25 | 1.25 | 2.50 | 1.25 | 1.25 | 1.25 | 1.25 |
| S-Ha-2 | 1.25 | 0 | 1.25 | 1.25 | 5 | 0 | 6.25 |
| D1-Ha-1 | 5 | 3.75 | 6.25 | 1.25 | 0 | 2.50 | 1.25 |
| D1-Ha-2 | 0 | 0 | 0 | 1.25 | 1.25 | 1.25 | 1.25 |
| D2-Ha-1 | 1.25 | 0 | 1.25 | 1.25 | 1.25 | 0 | 3.75 |
| D2-Ha-2 | 1.25 | 3.75 | 3.75 | 1.25 | 0 | 2.50 | 2.50 |

Following equation (1), ignoring the constants $\alpha^{(3)}$, $N$, and $P^{(\omega)}$, the E-FISH signal at a given displacement z, if the electric field extends to infinity is given by equation (C.1) below.

$$S(z) = \frac{1}{z_R}\left|\int_{-\infty}^{\infty} E_{ext}(z)\frac{e^{i\Delta kz}}{1+i\frac{z}{z_R}}dz\right|^2 \tag{C.1}$$

By differentiating equation (C.1) with respect to Rayleigh range, the sensitivity of the signal to Rayleigh range can be obtained, as shown in equation (C.2) below.

$$\frac{\partial S(z)}{\partial z_R} = -\frac{\left|\int_{-\infty}^{\infty} E_{ext}(z)\frac{e^{i\Delta kz}}{1+i\frac{z}{z_R}}dz\right|^2}{z_R^2} + \frac{2}{z_R}Re\left[\int_{-\infty}^{\infty} E_{ext}(z)\frac{e^{i\Delta kz}}{1+i\frac{z}{z_R}}dz^{*}\frac{\partial\left(\int_{-\infty}^{\infty} E_{ext}(z)\frac{e^{i\Delta kz}}{1+i\frac{z}{z_R}}dz\right)}{\partial z_R}\right]$$

$$\frac{\partial\left(\int_{-\infty}^{\infty} E_{ext}(z)\frac{e^{i\Delta kz}}{1+i\frac{z}{z_R}}dz\right)}{\partial z_R} = \int_{-\infty}^{\infty} E_{ext}(z)\frac{-z\sin(\Delta kz)+iz\cos(\Delta kz)}{z_R^2-z^2+2iz_Rz}dz \tag{C.2}$$

## Appendix D

Following equation (3), ignoring the constants $\alpha^{(3)}$, $N$, and $P^{(\omega)}$, the E-FISH signal at a given transverse location $x$, if the electric field extends to infinity is given by equation (D.1) below.

$$S(x) = \left|\int_{-\infty}^{\infty} E_{ext}\left(\sqrt{x^2+z^2}\right)\frac{e^{i\Delta kz}}{1+i\frac{z}{z_R}}dz\right|^2 \tag{D.1}$$

By differentiating equation (D.1) with respect to Rayleigh range, the sensitivity of the signal to Rayleigh range can be obtained, as shown in equation (D.2) below.

$$\frac{\partial S(x)}{\partial z_R} = Re\left[\int_{-\infty}^{\infty} E_{ext}\left(\sqrt{x^2+z^2}\right)\frac{e^{i\Delta kz}}{1+i\frac{z}{z_R}}dz^{*}\frac{\partial\left(\int_{-\infty}^{\infty} E_{ext}\left(\sqrt{x^2+z^2}\right)\frac{e^{i\Delta kz}}{1+i\frac{z}{z_R}}dz\right)}{\partial z_R}\right] \tag{D.2}$$

For the second term, given that $\sqrt{x^2+(-z)^2} = \sqrt{x^2+z^2}$, the integral limits can be shortened from 0 to ∞, with the multiplier term $\frac{e^{i\Delta kz}}{1+i\frac{z}{z_R}}$, split correspondingly for $z$ (labelled as $g_+(z)$) and $-z$ (labelled as $g_-(z)$), following equation (D.3) below.

$$g_+(z) = \frac{e^{i\Delta kz}}{1 + i\frac{z}{z_R}}$$

$$g_-(z) = \frac{e^{-i\Delta kz}}{1 - i\frac{z}{z_R}}$$

(D.3)

Using the properties for complex numbers, the following simplified relationship for the multiplier term is given by equation (D.4) below. The *Re* in equation (D.4) indicates that only the real portion of the signal is taken in.

$$g_+(z) + g_-(z) = 2\,Re(g_+)$$
$$Re(g_+) = \frac{z_R^2\cos(\Delta kz) + zz_R\sin(\Delta kz)}{z_R^2 + z^2}$$

(D.4)

Substituting equation (D.4), equation (D.1) simplifies to equation (D.5) below.

$$S(x) = 2\int_0^\infty E_{ext}\left(\sqrt{x^2+z^2}\right)\frac{z_R^2\cos(\Delta kz) + zz_R\sin(\Delta kz)}{z_R^2 + z^2}dz$$

(D.5)

As $z_R$ only appears in the multiplier term, we differentiate the multiplier term (labelled as $K(z, z_R)$) with respect to $z_R$ to assess its sensitivity, as given by equation (D.6) below.

$$\frac{\partial K(z,z_R)}{\partial z_R} = \frac{2z_Rz^2\cos(\Delta kz) - z(z_R^2 - z^2)\sin(\Delta kz)}{(z_R^2+z^2)^2}$$

(D.6)

Thus, the sensitivity of the signal is given by equation (D.7) below.

$$\frac{\partial S(x,z_R)}{\partial z_R} = Re\left[\begin{array}{c}\left(\int_{-\infty}^{\infty} E_{ext}\left(\sqrt{x^2+z^2}\right)\frac{e^{i\Delta kz}}{1+i\frac{z}{z_R}}dz\right)^* \cdot \\ \int_0^\infty 2E_{ext}\left(\sqrt{x^2+z^2}\right)\frac{2z_Rz^2\cos(\Delta kz) - z(z_R^2-z^2)\sin(\Delta kz)}{(z_R^2+z^2)^2}dz\end{array}\right]$$

(D.7)